\documentclass{aa}  

\usepackage{graphicx}
\usepackage{txfonts}
\newcommand{\msun}{{\rm M_{\sun }}}
\newcommand{\Mo}{\msun}
\begin{document}

   \title{Prograde and retrograde stars in nuclear cluster mergers}
   \subtitle{Evolution of the supermassive black hole binary and the host galactic nucleus}

   \author{A. Mastrobuono-Battisti
          \inst{1,2}
     \and P. Amaro Seoane\inst{3,4,5}
     \and M. J. Fullana i Alfonso\inst{3}
     \and C. Omarov \inst{6}
    \and D. Yurin \inst{6}
    \and M. Makukov \inst{6}
    \and G. Omarova \inst{6}
    \and G. Ogiya \inst{7}
    }

   \institute{         Dipartimento di Fisica e Astronomia ``Galileo Galilei'', Univ. di Padova, Vicolo dell’Osservatorio 3, Padova 35122, Italy\\
              \email{alessandra.mastrobuono@unipd.it}
         \and 
         GEPI, Observatoire de Paris, PSL Research University, CNRS, Place Jules Janssen, 92190 Meudon, France
         \and
          Institut de Matemàtica Multidisciplinària. Universitat Politècnica de València, Camí de Vera, s/n, 46022, València, Spain
         \and
          Max Planck Institute for Extraterrestrial Physics, Garching, Germany
         \and
          Higgs Centre for Theoretical Physics, Edinburgh, UK
         \and
            Fesenkov Astrophysical Institute 050020 Almaty, Kazakhstan
            \and
            Institute for Astronomy, School of Physics, Zhejiang University, Hangzhou 310027, China\\
             }

   \date{\today}

 
  \abstract
   {Nuclear star cluster (NSC) mergers, involving the fusion of dense stellar clusters near the centres of galaxies, play a pivotal role in shaping galactic structures. The distribution of stellar orbits has significant effects on the formation and characteristics of extreme mass ratio inspirals (EMRIs).} %
   {In this study, we address the orbital distribution of stars in merging NSCs and the subsequent effects on supermassive black hole binary (SMBHB) evolution. } %
   {We ran dedicated direct-summation $N$-body simulations with different initial conditions to do a detailed study of the resulting NSC after their progenitors had merged.} %
   {Our findings reveal that prograde stars form a flattened structure, while retrograde stars have a more spherical distribution. The axial ratios of the prograde component vary based on the presence and mass ratio of the SMBHs. The fraction of prograde and retrograde stars depends on the merger orbital properties and the SMBH mass ratio. The interactions of retrograde stars with the SMBHB affect the eccentricity and separation evolution of the binary.  Our analysis reveals a strong correlation between the angular momentum and eccentricity of the SMBH binary. This relationship could serve as a means to infer information about the stellar dynamics surrounding the binary. We find that prograde orbits are particularly close to the binary of SMBHs, a promising fact regarding EMRI production. Moreover, prograde and retrograde stars have different kinematic structures, with the prograde stars typically rotating faster than the retrograde ones. The line-of-sight velocity and velocity dispersion, as well as the velocity anisotropy of each NSC, depend on the initial merger orbital properties and SMBH mass ratios. The prograde and retrograde stars always show different behaviours.} %
   {The distribution of stellar orbits and the dynamical properties of each kinematic population can potentially be used as a way to tell the properties of the parent nuclei apart, and has an important impact on expected rates of EMRIs, which will be detected by future gravitational wave observatories such as the Laser Interferometer Space Antenna (LISA).}
               
\maketitle
%

\section{Introduction}
Nuclear star clusters (NSCs) are dense and compact stellar systems hosted by galaxies with masses typically between $10^8\,\Mo$ and $10^{10}\,\Mo$ \citep{Boeker04, Cote06, Boeker10, Neumayer11, Turner12, Georgiev14, denbrok14, SJ19, Neumayer20}. NSC properties are linked to those of their host \citep{Rossa06, Ferrarese06, Wehner06, Neumayer20}, and their formation process is thought to be the result of in situ star formation \citep{Loose82, LE03, Milosavljevic04, Nayakshin05, Paumard06, Schinnerer06, Schinnerer08, Hobbs09, Mapelli12, Mastrobuono19} and dynamical friction-driven star cluster decay and mergers \citep{Tremaine75, CD93, Antonini12, Gnedin14, Mastrobuono14, PMB14, Arcasedda15, AN15, Tsatsi17, Abbate18}. NSCs can coexist with the central supermassive black hole (SMBH), as in the case of the Milky Way, which hosts Sgr A$^*$, an SMBH of $4.3\times10^6\,\Mo$, at the centre of an NSC of about $2.5\times10^{7}\,\Mo$ and with a half-light radius of about $4$\,pc \citep{SchoedelEtAl03, EI05, GhezEtAl05, GhezEtAl08, GillessenEtAl09, BGS16, GP17, Schoedel14b}.\\

Nuclear star cluster mergers, involving the coalescence of dense stellar clusters near the centres of galaxies, play a crucial role in the formation and evolution of galactic structures. Understanding the dynamics of these mergers, particularly the distribution of stellar orbits, is of great importance for unravelling the underlying mechanisms and their broader astrophysical implications.

The distribution of stellar orbits during the merger of NSCs has several significant implications. Firstly, it directly affects the structural evolution of the merging system. The orbital characteristics of stars within the NSCs determine their interactions and subsequent dynamical evolution. The distribution of stellar orbits influences the overall density profile, shape, and kinematics of the merged cluster, thereby shaping the resultant galactic nucleus and its surrounding structures.

Secondly, the distribution of stellar orbits in NSC mergers has implications on the growth and activity of central SMBHs. SMBHs are commonly found in galactic nuclei \citep[see e.g.][]{Ferrarese05, Kormendy13, Neumayer12, Nguyen19, Neumayer20}, and their interactions with the surrounding stellar population significantly influence the stellar orbits. Stellar encounters with SMBHs can alter the distribution of prograde and retrograde orbits, affecting the stellar density near the SMBH and potentially driving accretion events \citep{Sesana11, Gualandris12_2}.

Studying the distribution of stellar orbits is also crucial for understanding the formation of various astrophysical structures. During NSC mergers, the complex interplay of gravitational interactions leads to the formation of stellar discs, bars, and other non-axisymmetric structures. The distribution of stellar orbits directly affects the formation and properties of these structures, which in turn influence the dynamical evolution of galaxies and their observable features \citep[see e.g.][]{MM01}.

Furthermore, the distribution of stellar orbits in NSC mergers is closely connected to the production of gravitational waves \citep[GWs, see e.g.][]{Gourgoulhon19}. The presence of binary systems and the orbital interactions of stars contribute to the emission of GWs, which can be detected by GW observatories. Understanding the distribution of stellar orbits and their properties is essential for predicting and interpreting the GW signals originating from NSC mergers.

By studying the distribution of stellar orbits in the mergers of NSCs, we gain valuable insights into the mechanisms driving galactic evolution, black hole growth, and the production of GWs. This knowledge helps us understand the formation and evolution of galactic structures, the dynamics of central SMBHs, and the astrophysical processes involved in the emission of GWs.
Understanding the distribution of prograde and retrograde stellar orbits in the mergers of NSCs is of paramount importance, particularly in the context of extreme mass ratio inspirals \citep[EMRIs, see e.g.][]{Amaro-SeoaneLRR2012,2022hgwa.bookE..17A,Amaro-SeoaneEtAl07}. EMRIs involve the inspiral of a stellar-mass compact object, such as a white dwarf or a stellar black hole, into a much more massive SMBH at the centre of a galaxy.

The distribution of prograde and retrograde stellar orbits plays a crucial role in determining the formation and characteristics of EMRIs. Specifically, the location of the last stable orbit, known as the innermost stable circular orbit (ISCO), depends on the spin of the SMBH and the orientation of the orbit relative to the black hole's spin axis \citep{Amaro-Seoane2013}. Prograde orbits, where the angular momentum of the stellar object aligns with the black hole's spin, tend to have larger ISCO radii; whereas, retrograde orbits, with angular momentum in the opposite direction, result in smaller ISCO radii.
The significance of understanding this distribution lies in its direct impact on the event rate of EMRIs. Prograde orbits contribute to EMRI events at greater distances from the SMBH, leading to an increased event rate per unit volume. On the other hand, retrograde orbits, while still contributing to the event rate, have a smaller effect compared to prograde orbits, as they lead to EMRIs occurring at closer distances to the SMBH \citep{Amaro-Seoane2013}.
This information is particularly crucial for future space-based GW observatories such as the Laser Interferometer Space Antenna \citep[LISA,][]{Amaro-SeoaneEtAl2017}. LISA aims to detect and study low-frequency GWs, including EMRIs. The event rate and characteristics of EMRIs observed by LISA strongly depend on the distribution of prograde and retrograde stellar orbits in NSC mergers. By understanding this distribution, we can better estimate the expected rates of EMRIs and plan observational strategies accordingly.
Hence, comprehending the distribution of prograde and retrograde stellar orbits in NSC mergers is essential for understanding the formation of EMRIs. The location of the last stable orbit, influenced by the orientation of the orbit and the SMBH's spin, directly affects the event rate of EMRIs. Prograde orbits contribute to larger distances, resulting in increased event rates, while retrograde orbits lead to smaller distances, diminishing the event rate, although to a lesser extent compared to prograde orbits. This knowledge is crucial for future GW observatories such as LISA, as it helps inform the expected rates and characteristics of EMRI events.

In this paper, we address the evolution of the stellar orbital distribution during the NSC merger, as a function of the merger initial conditions and of the NSC and SMBH dynamical and structural properties. By employing numerical simulations, we analyse the underlying mechanisms shaping the observed orbital distributions. Our study contributes to the broader understanding of NSC mergers, their impact on galactic evolution, and the formation of astrophysical structures.
We have done so by running new simulations, similar to those presented by \cite{Ogiya20} and \cite{Mastrobuono23}, to explore the presence of retrograde stars in cores surrounding an SMBHB, a single SMBH, or no black hole, after a merger of two nucleated galaxies. 

   \begin{figure}
   \raggedright
   \includegraphics[width=\columnwidth]{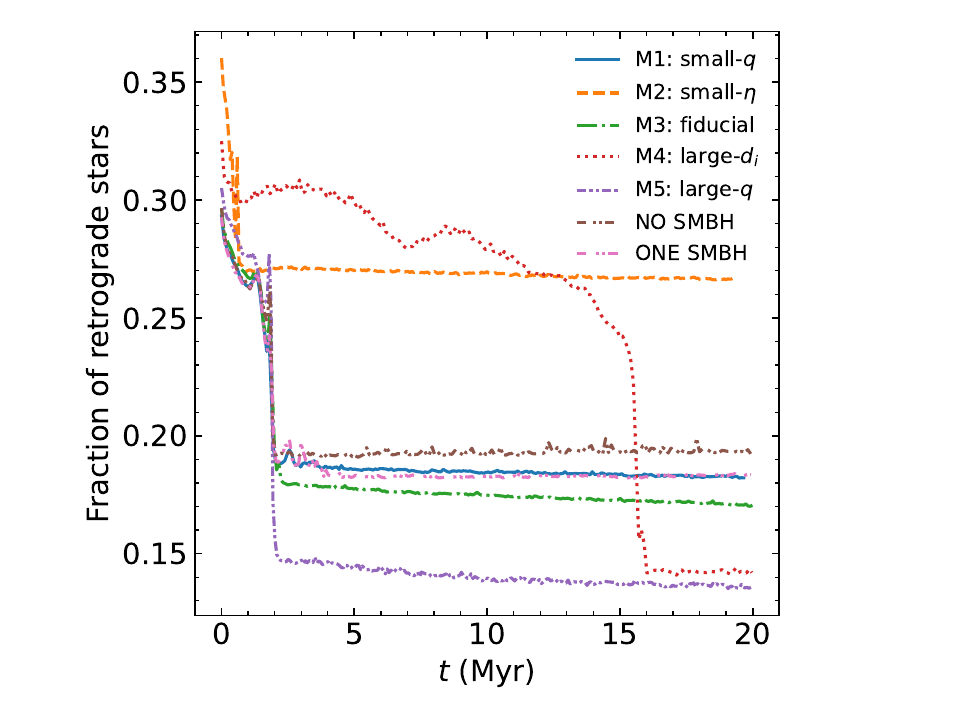}
    \includegraphics[width=\columnwidth]{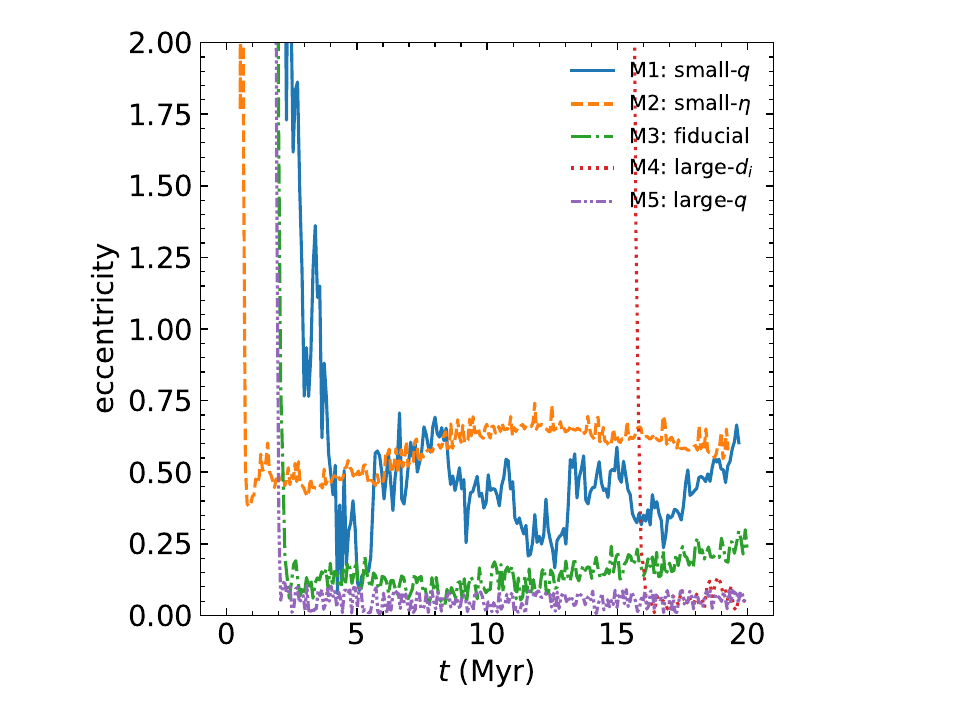}
    \includegraphics[width=\columnwidth]{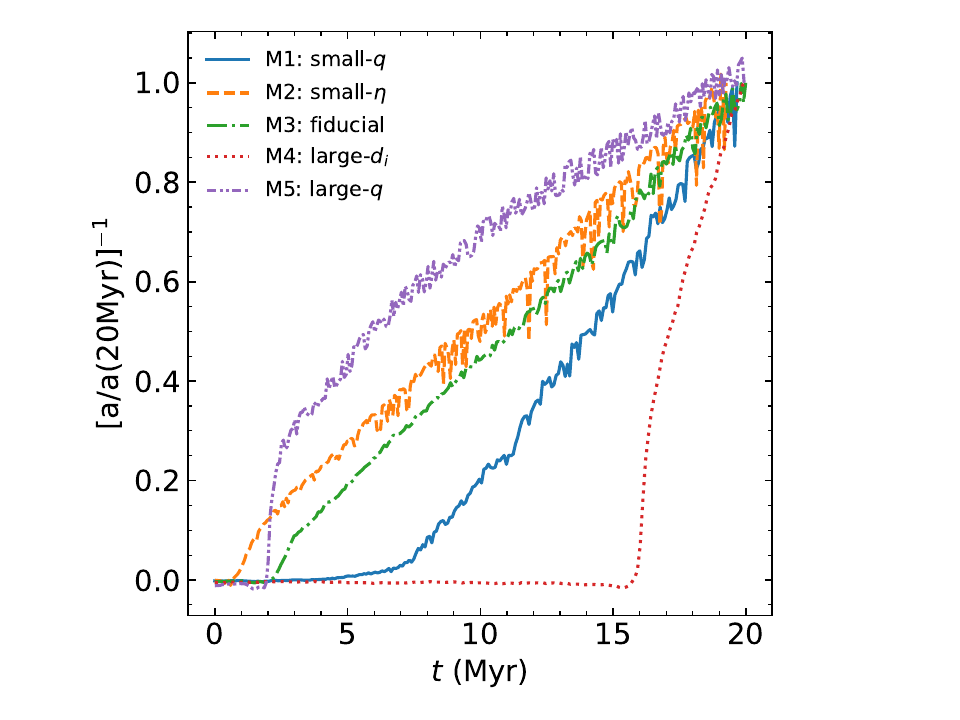}
      \caption{Fraction of retrograde stars (top panel), the SMBHB eccentricity (middle panel), and the inverse of the binary separation rescaled to the final value (bottom panel) as a function of the time for the five NSC merger simulations containing two SMBHs.
              }
         \label{fig:fract_retrograde}
   \end{figure}
\begin{table*}
    \caption{Initial and final properties of the simulated models.}
    \centering
        \begin{tabular}{c c c c c c c c c c c c}        
        \hline\hline 
        ID: property & $q$ ($M_{\bullet}/M_\odot$) & $d_i$ (pc) & $\eta$ & $f_{r, i}$ & $f_{r, f}$ & $f_{r1, f}$ & $f_{p1, f}$ & $M_{r}\,(\Mo)$ & $r_{h,p}$ (pc) & $r_{h,r}$ (pc) & $r_{h,tot}$ (pc) \\
        \hline
        M1: small-$q$ & 0.01 & 20 & 1.0 & 0.29 & 0.18 & 0.74 & 0.45 & $3.5\times10^6$ & 10 & 4.6 & 8.6 \\
        M2: small-$\eta$ & 0.1 & 20 & 0.5 & 0.36 & 0.27 & 0.60 & 0.46 & $5.1\times10^6$ & 9.0  & 7.3 & 8.6 \\
        M3: fiducial & 0.1 & 20 & 1.0 & 0.29 & 0.17 & 0.73 & 0.44 & $2.7\times10^6$ &  11 & 5.5 & 9.8 \\
        M4: large-$d_i$ & 0.1 & 50 & 1.0 & 0.33 & 0.14 & 0.78 & 0.45 & $2.7\times10^6$ & 11 & 3.9 & 9.3 \\
        M5: large-$q$ & 1.0 & 20 & 1.0 & 0.31 & 0.14 & 0.50 & 0.50 & $2.4\times10^6$ & 13 & 7.4 & 12\\
        M6: no-SMBH  & (0)& 20 & 1.0 & 0.30 & 0.19 & 0.49 & 0.50 & $3.8\times10^6$ & 9.6 & 4.0  & 8.0\\
        M7: one-SMBH & ($10^6$) &  20 & 1.0  & 0.29 & 0.18 & 0.74 & 0.45 & $3.6\times10^6$ & 9.7 & 4.0  & 8.2\\
         \hline
    \end{tabular}
    \tablefoot{The table lists the name and main defining property of each model (ID: property), the SMBH mass ratio ($q$) or the mass of the single SMBH, if present ($M_{\bullet}/M_\odot$), the initial distance $d_i$ between the SMBHs (or between the centres of mass of the systems in the case of models with only one or no SMBH), the parameter $\eta$ which quantifies the initial relative velocity between the NSCs, the initial and final fraction of retrograde stars, $f_{r, i}$ and $f_{r, f}$, the fraction of retrograde and retrograde stars coming from the primary NSC ($f_{r1, f}$ and $f_{p1, f}$), the total mass in retrograde stars ($M_{r}$),  the half-mass radius of the prograde ($r_{h,p}$) and retrograde ($r_{h,r}$) components, as well as that of the entire system ($r_{h,tot}$).}
    \label{tab:tab1}
\end{table*}
\begin{table}
    \caption{Properties of the SMBHBs at the end of the simulations.}
    \centering
    \setlength{\tabcolsep}{4pt}  
        \begin{tabular}{@{}c c c c c@{}}  
        \hline\hline 
        ID: property & $e_f$ (pc) & $a_f$ (pc) & $d_{min, p}$ (pc) & $d_{min, r}$ (pc)  \\
        \hline
        M1: small-$q$ & 0.60 & $1.80\times10^{-3}$ & $1.7\times10^{-2}$ & $3.8\times10^{-2}$ \\
        M2: small-$\eta$ & 0.58 & $3.09\times10^{-3}$ & $3.7\times10^{-2}$ & $3.5\times10^{-3}$\\
        M3: fiducial & 0.22 & $3.37\times10^{-3}$ & $1.8\times10^{-2}$ & $3.4\times10^{-2}$\\
        M4: large-$d_i$ &  0.070 & $1.20\times10^{-2}$ & $4.3\times10^{-3}$ & $7.6\times10^{-3}$\\
        M5: large-$q$ &  0.055 & $2.31\times10^{-2}$ & $1.1\times10^{-1}$ & $1.4\times10^{-1}$\\
        \hline
    \end{tabular}
    \tablefoot{$e_f$ is the final eccentricity of the SMBHB, $a_f$ is its final separation,  $d_{min, p}$ and $d_{min, r}$ are the distances of the closest prograde and retrograde star to the centre of the final NSC, respectively. }
    \label{tab:tab2}
\end{table}

\section{Models and methods}
We analyse a new set of collisional simulations with similar initial conditions as those presented in  \cite{Ogiya20} and \cite{Mastrobuono23}.
In such simulations, two NSCs merge to form a new galactic nucleus that contains an SMBHB, a single SMBH or no SMBH. We used our models to perform a taxonomical study of the resulting NSC depending on the combining properties and the composing stars' kinematic and orbital properties.
All the simulations were performed with NBODY6++GPU \citep{Wang15}, an updated version of the widely used direct $N$-body code NBODY6 \citep{Aarseth03}, improved to run on multiple graphic processing units.  All the input parameters necessary to reproduce the simulations are described in \cite{Ogiya20} for the models with two SMBHs and \cite{Mastrobuono23} for all models, including those with only one or no SMBH. In the next paragraphs, we summarize the simulation setup and describe the models.
As the models and the simulations are described in detail in \cite{Ogiya20} and \cite{Mastrobuono23}, here we briefly summarize their properties. 
The density profile of the NSCs is modelled as a \cite{Dehnen93} profile
\begin{equation}
    \rho(r)=\frac{(3-\gamma)M_{NSC}}{4\pi}\frac{r_0}{r^\gamma(r+r_0)^{4-\gamma}}
,\end{equation}
where $M_{NSC}$ is the total NSC mass, $r_0$ is its core radius and $\gamma$ is the inner slope of the NSC density profile.
In all our models, the initial total mass of each NSC is $M_{NSC}=10^7\,\Mo$. All the simulated NSCs have a cored density profile with $\gamma=0$ and a core radius $r_0=1.4\,$pc, which corresponds to a half-light radius of $4\,$pc, a value typical for NSCs of mass similar to the adopted one \citep{Neumayer20}. In all simulations with two SMBHs (i.e. models M1 to M5), the primary NSC hosts an SMBH of
$10^6\,\Mo$ while the secondary NSC hosts either a $10^4\,\Mo$, $10^5\,\Mo$ or $10^6\,\Mo$ SMBH. Those mass values span the SMHB mass range observed in NSC of about $10^7\,\Mo$  \citep{Georgiev16}. We  modelled each NSC using 65\,536 single-mass $N$-body particles, leading to a mass resolution of about $152.6\,\Mo$. We introduced the central SMBH with zero velocity. The velocities of the stellar particles are drawn using the \cite{Eddington16} formula, in which the SMBH is taken into account when calculating the gravitational potential. Model M6 contains no central SMBH, while in M7 only one of the two merging NSC hosts an SMBH of $10^6\,\Mo$.\\
The initial separation between the centres of the two NSCs, $d_i$, is either $20\,$pc or $50\,$pc. The two SMBHs are initially unbound. We define the parameter $\eta$ which sets the initial relative velocity between the NSCs 
\begin{equation}
    v_i = \eta\sqrt{\frac{GM_*(d_i)}{d_i}}
,\end{equation}
where $M_*(d_i)$ is the mass of the merging systems calculated as the sum of the NSC masses enclosed within a distance $d_i/2$ from their centres. In our models, $\eta$ is either equal to 0.5 or 1. 
Our NSCs are on circular relative orbits, following the expected orbital circularization \citep{Penarubia2004}, and have parallel orbital angular momentum.  The position of the primary SMBH is initially at the origin of the reference frame, where it is located with zero velocity. The secondary SMBH is initially located on the $x$-axis with a total velocity $v_i$, oriented in the $y$ direction. The initial parameter set-up for each model, along with the names adopted for them, are summarized in Table \ref{tab:tab1}. 


\section{Results}
We analyse the time evolution of the orbital distribution of the stars in the progenitors and the final NSC as a function of the merger properties and dynamical and structural characteristics of the systems. We focus in particular on the redistribution of the stars in two families, those that co-rotate with the orbital angular momentum of the merging NSCs (prograde stars) and those that counter-rotate with it (retrograde stars). This redistribution around the SMBHB is crucial to determine its evolution and the kind of GW signal emitted by EMRIs and also to determine the evolution of the SMBH binary and the kind of GW signal emitted during the black hole coalescence.
In our analysis, we only consider bound stars (i.e. stars with negative energy with respect to the centre of density of the system) and we move to the reference frame in which the total angular momentum of the system is aligned with the z-axis. 
In this reference frame, prograde (retrograde) stars have an angular momentum of the same (opposite) sign of the total angular momentum of the system. The total angular momentum of a star is the combination of the cluster stellar angular momentum with respect to the centre of the host NSC and the orbital angular momentum of the merging system.
\subsection{Time evolution of the retrograde stellar fraction}
The two progenitor NSCs show no internal rotation at the beginning of the simulation, therefore, 
their stars are distributed on orbits with randomly oriented angular momenta. When the two NSCs are put on their relative orbit, a fraction of the stars will co-rotate with the NSCs' orbital spin and the rest will counter-rotate, determining the initial orbital distribution of the system.
During the merger, the ongoing dynamical interactions, modify the orbits of the stars, affecting the orientation of their angular momenta and leading to a different final orbital distribution.
To follow this effect, we study how the fraction of retrograde stars changes with time, depending on the initial model configuration. 
Table \ref{tab:tab1} shows the initial and final overall retrograde stellar fractions and total stellar masses, as well as the retrograde and prograde fractions of stars coming from the primary NSC. All models start with a similar fraction of retrograde stars. This fraction is slightly larger for models M2 and M4, which have a different initial orbital angular momentum with respect to the other systems.
In all models, the fraction of retrograde stars decreases rapidly during the merger, to then approximately plateau around the final value when the merger is complete (see top panel of Fig. \ref{fig:fract_retrograde}). Only in the case of M4, which has a larger initial distance compared to the other models, the value has a relatively small decrease at around 2~Myr, and then a sharper decrease at around 15~Myr when the merger is complete\footnote{In all other cases, the merger is complete after less than 2~Myr.}. M4, which has an initial larger distance between the two merging systems, and M5, which is characterized by the largest SMBH mass ratio ($q=1$), show the lowest final retrograde stellar fractions ($0.14$).  
Model M2, which has the smallest amount of initial angular momentum and the most significant initial fraction of retrograde stars, shows the largest fraction of counter-rotating stars at the end of the simulation ($0.27$, which is approximately $1.5$ times that of the other models).
The rest of the models have a similar final fraction of retrograde stars, ranging between $0.17$ and $0.19$. Interestingly, M7, the model with only one SMBH, behaves similarly to M1, which, on the other hand, is characterized by the smallest mass ratio between the two SMBHs. The two progenitor NSCs contribute almost equally to the fraction of prograde stars, while their contribution to the fraction of retrograde stars is remarkably different, as seen from Table \ref{tab:tab1}. Only in the cases of a large $q$ and no SMBH, the two NSCs supply the same fraction of retrograde stars to the final NSC. In the case of a small initial orbital momentum (M2), there is a slight predominance of retrograde stars coming from the primary NSC. In the remaining models, approximately 75\% of the retrograde stars originate from the primary NSC. This suggests that retrograde stars that originate from the primary NSC are more likely to reach the central region of the final NSC while remaining bound to their SMBH and maintaining their initial angular momentum orientation. 

\begin{figure}
\includegraphics[trim=0 0 5mm 0,clip,width=0.505\columnwidth]{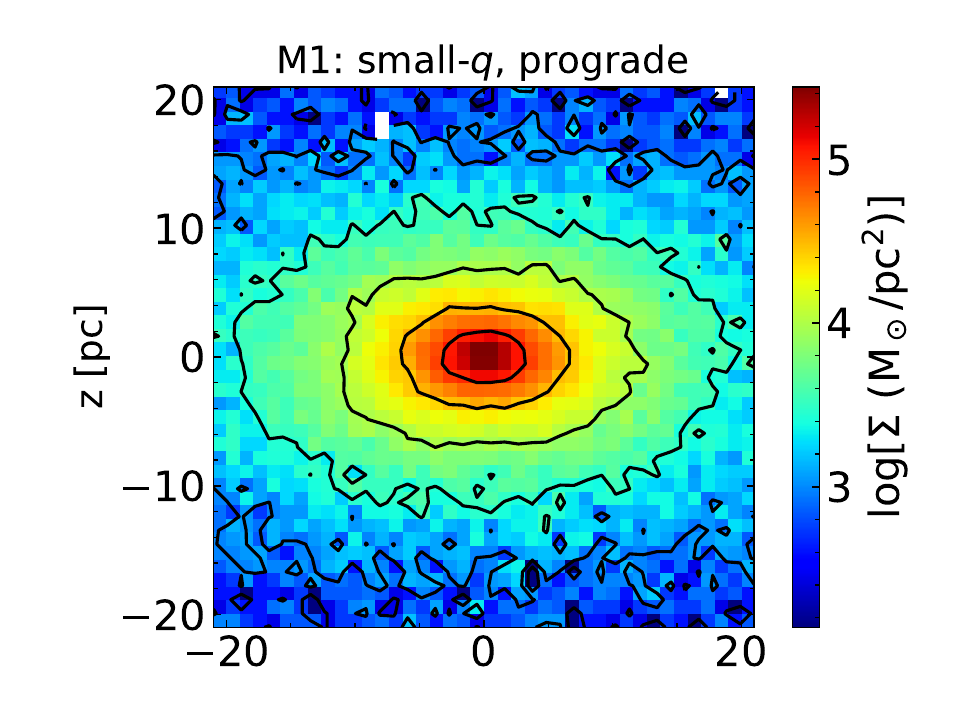}~\includegraphics[trim=0 0 5mm 0,clip,width=0.5\columnwidth]{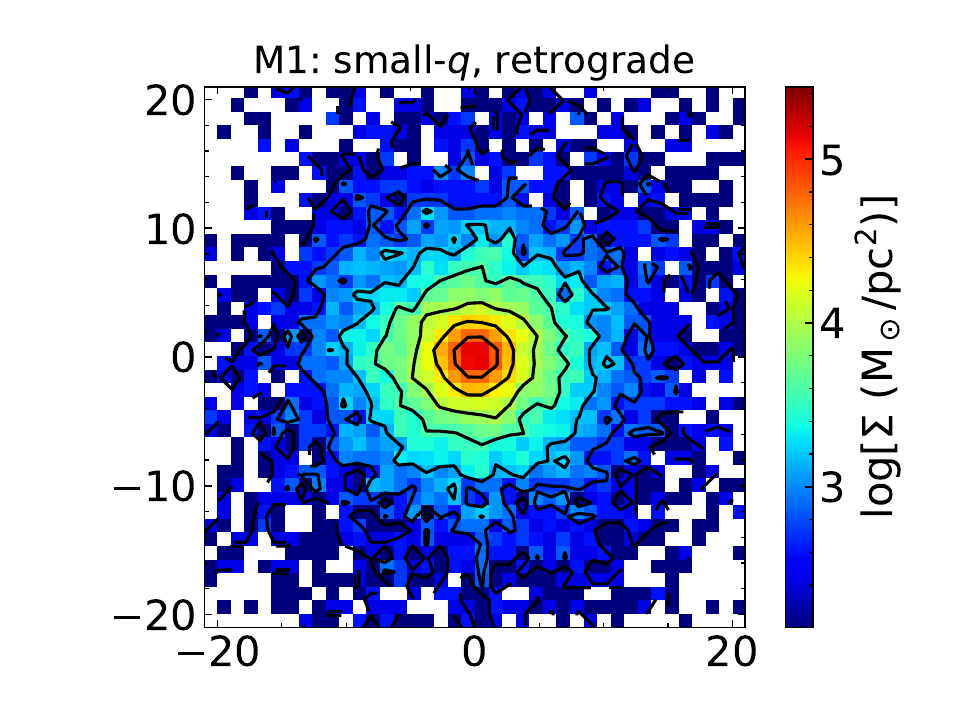}\\
\includegraphics[trim=0 0 5mm 0,clip,width=0.505\columnwidth]{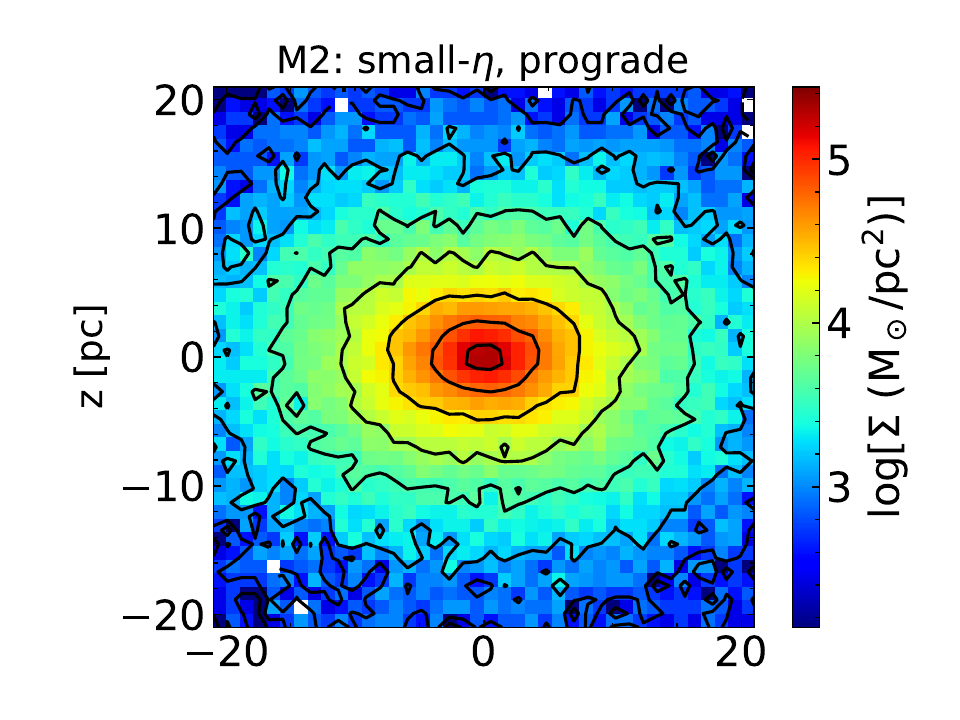}~\includegraphics[trim=0 0 5mm 0,clip,width=0.5\columnwidth]{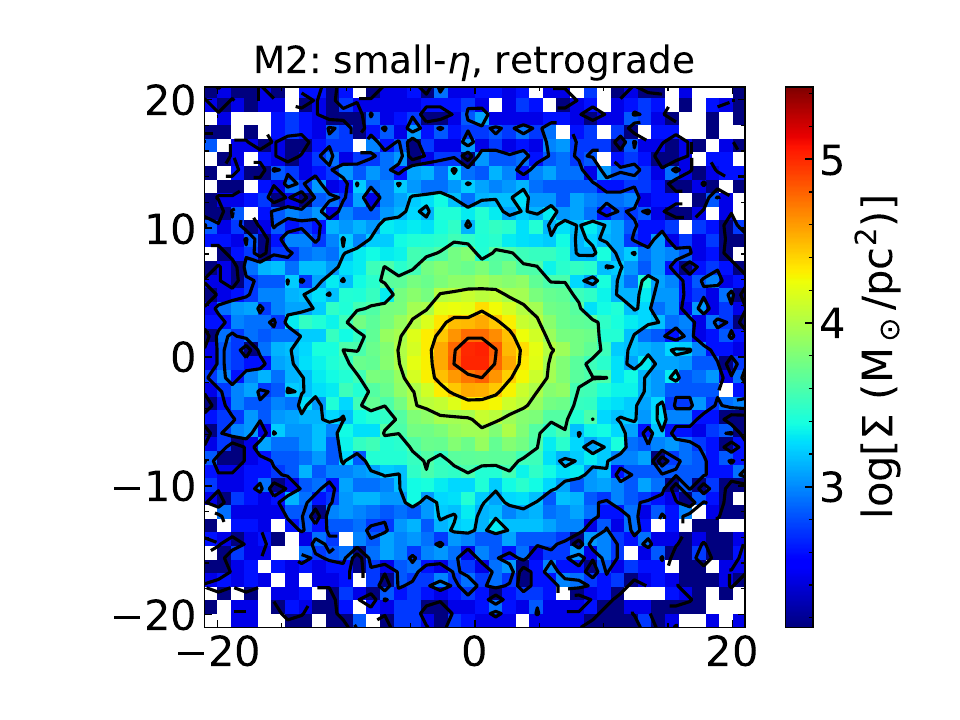}\\
\includegraphics[trim=0 0 5mm 0,clip,width=0.505\columnwidth]{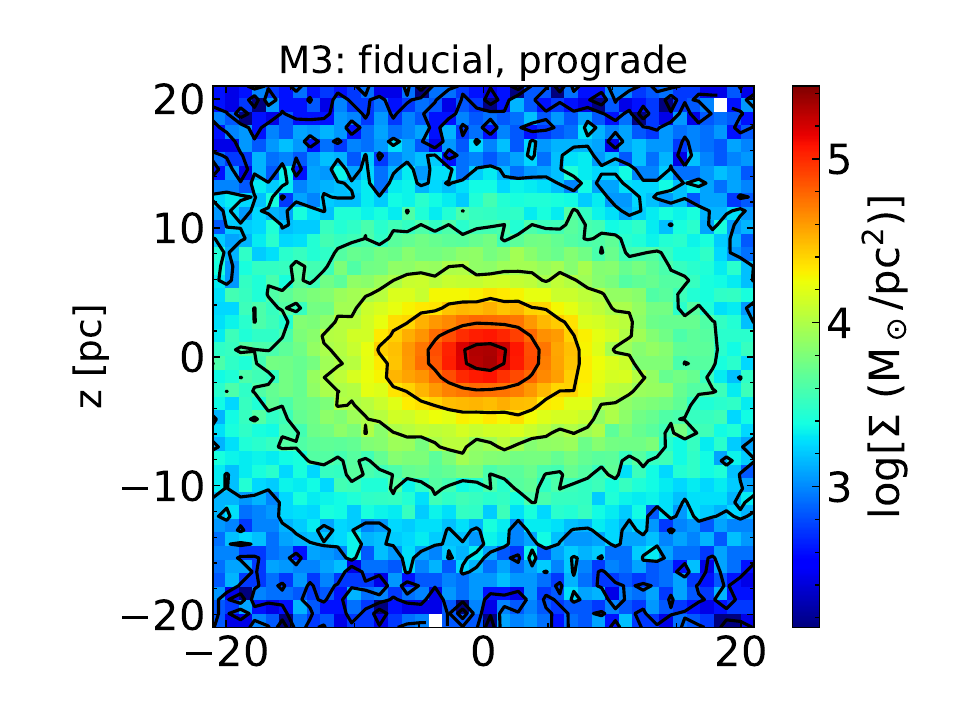}~\includegraphics[trim=0 0 5mm 0,clip,width=0.5\columnwidth]{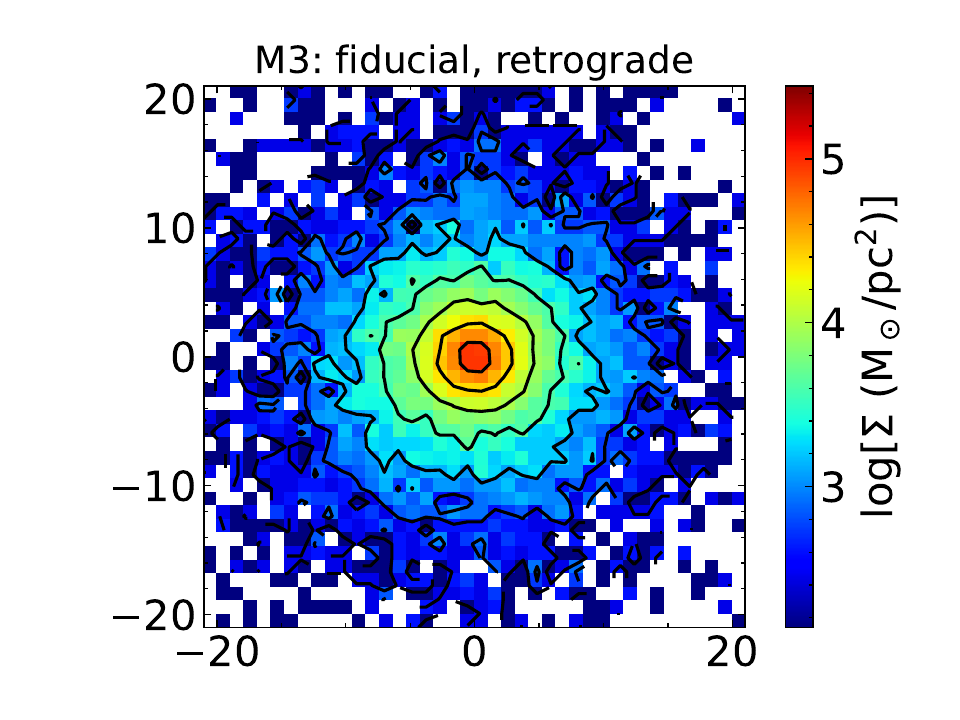}\\
\includegraphics[trim=0 0 5mm 0,clip,width=0.505\columnwidth]{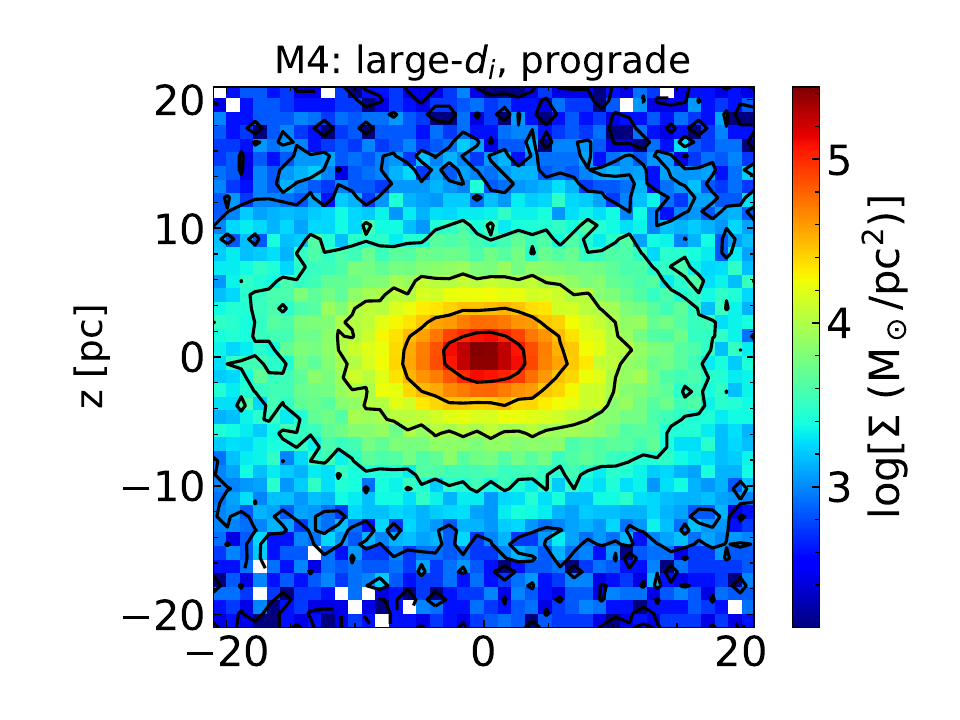}~\includegraphics[trim=0 0 5mm 0,clip,width=0.5\columnwidth]{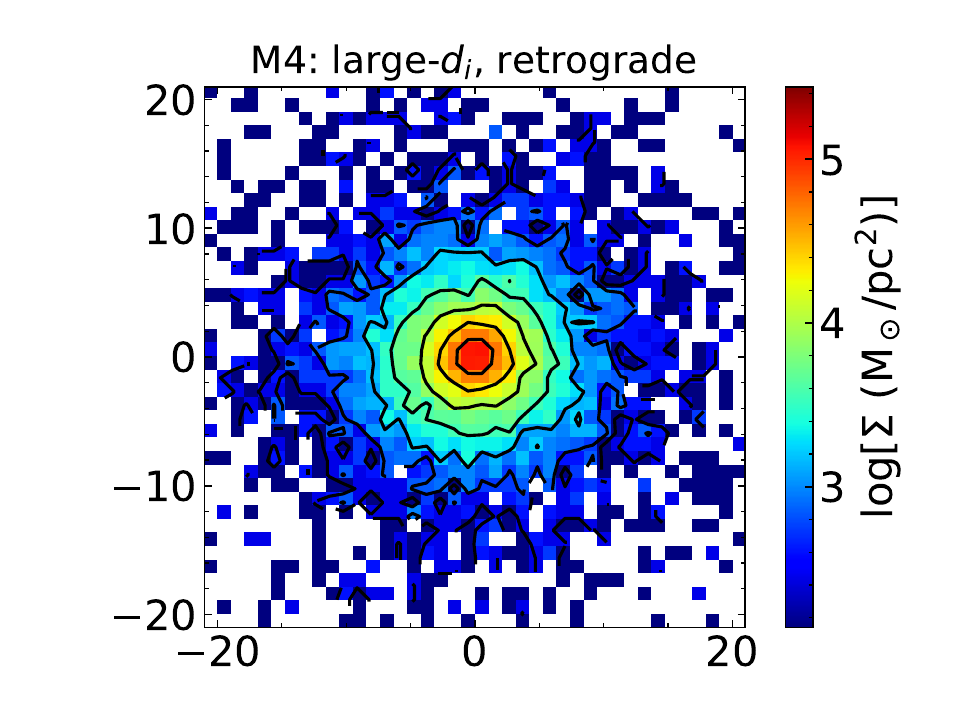}\\
\includegraphics[trim=0 0 5mm 0,clip,width=0.505\columnwidth]{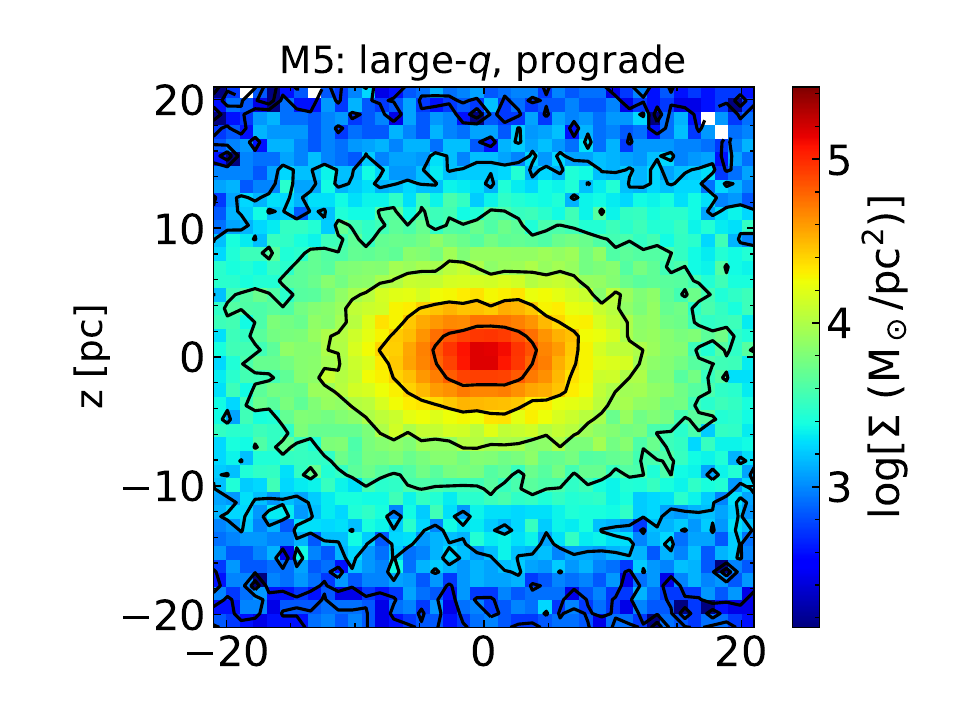}~\includegraphics[trim=0 0 5mm 0,clip,width=0.5\columnwidth]{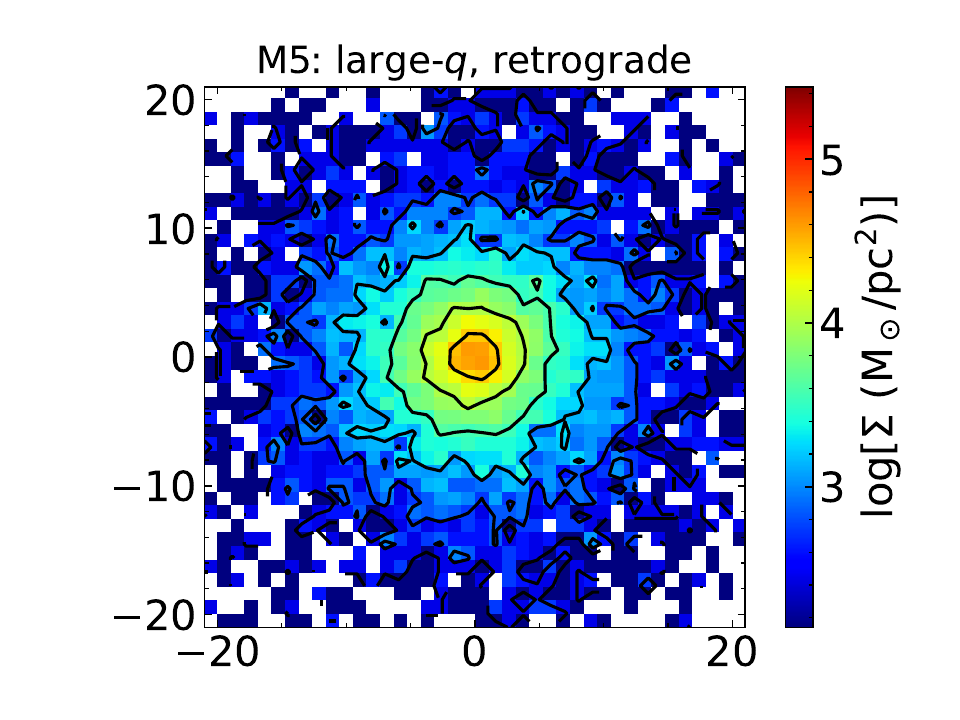}\\
\includegraphics[trim=0 0 5mm 0,clip,width=0.505\columnwidth]{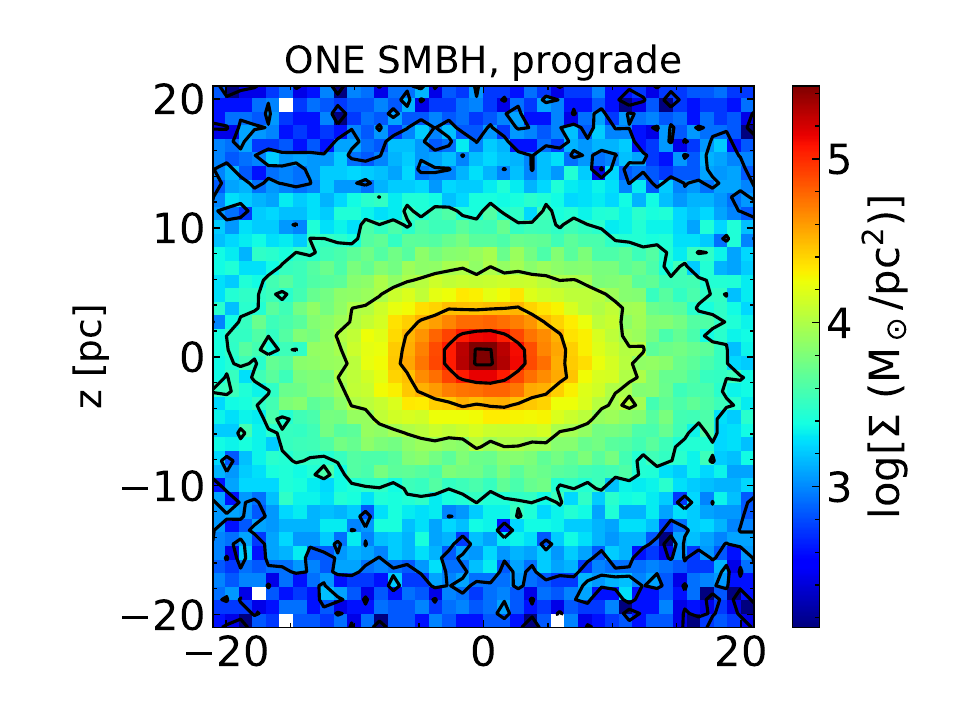}~\includegraphics[trim=7mm 0 5mm 0,clip,width=0.5\columnwidth]{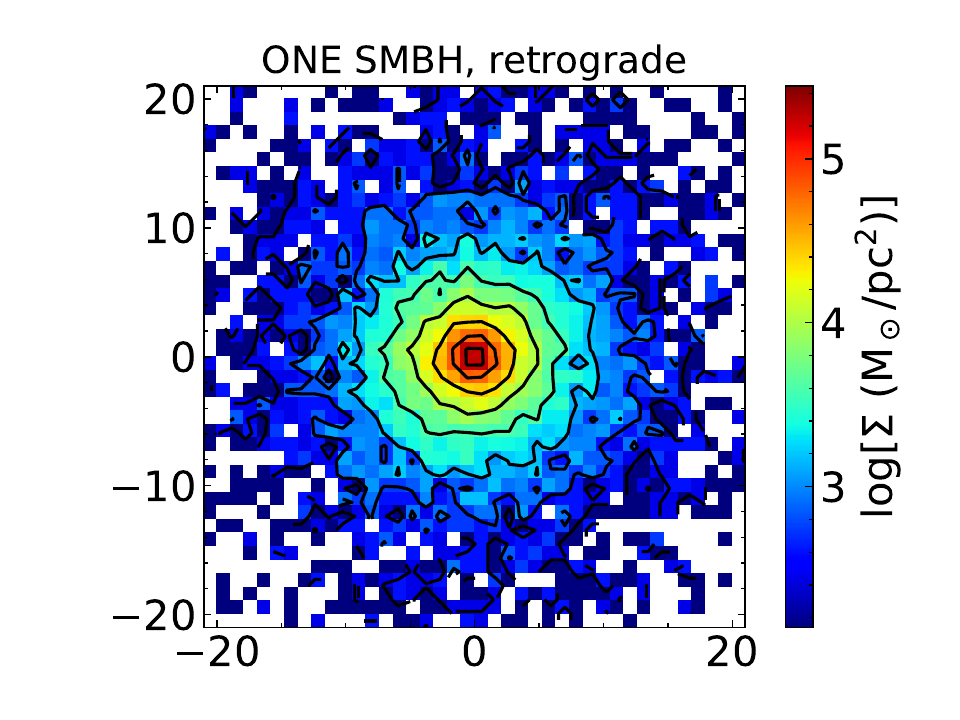}\\
\includegraphics[trim=11mm 0 5mm 0mm,clip,width=0.53\columnwidth]{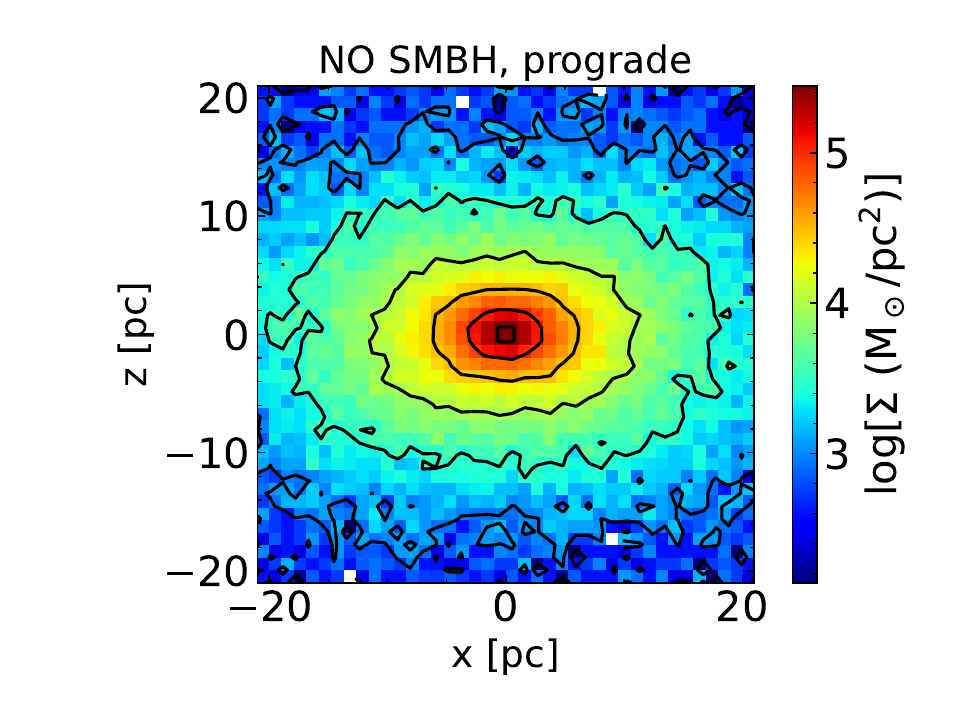}~\includegraphics[trim=20mm 0 10mm 0mm,clip,width=0.49\columnwidth]{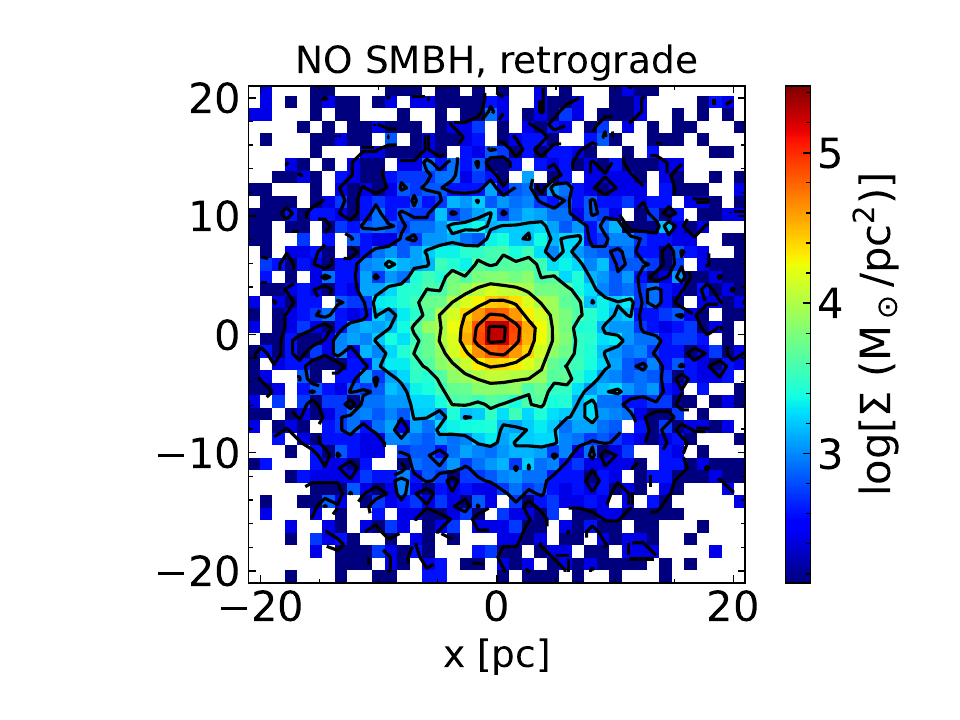}
      \caption{Density maps of the prograde (left panels) and retrograde (right panels) populations for each of the simulated models. The retrograde population is always more centrally concentrated and spherical than the prograde stellar component.
              }
         \label{fig:den_maps}
   \end{figure}
\subsection{The eccentricity and separation of the SMBHB}
The middle and lower panels in Fig. \ref{fig:fract_retrograde} display the evolution of the SMBHB's eccentricity and semi-major axis over time. The eccentricity rapidly decreases during the merger phase and exhibits varying behaviour afterwards, depending on the initial orbital configuration of the system.
The low $q$ binary formed in M1 exhibits significant eccentricity oscillations, ultimately reaching a final value of $0.6$ (see Table \ref{tab:tab2}), the highest among our models. We note that, possibly due to the different initial conditions, we reach a higher final eccentricity compared to what was found by \citep{Ogiya20} for the same model. This difference is due to the large oscillations of the eccentricity value that characterize this model. 

M2 shows a similar final value of the binary eccentricity compared to M1, however, the evolution of such quantity is smoother. After the initial sharp decrease caused by the SMBHs becoming bound, the eccentricity experiences a minor increase followed by a smaller decrease.
In our fiducial model, M3, initially, the eccentricity quickly decreases, reaching a value of about 0.10 when the SMBHs become bound. From 10~Myr on, the eccentricity shows a growing trend, with a final value of 0.22.  Models M4 and M5 show the lowest final SMBHB eccentricity values (0.0070 and 0.0055 respectively) among our models. This suggests that a larger amount of lost angular momentum or a higher SMBH mass ratio leads to low final binary eccentricities. In addition, we note that in M4 and M5, the fraction of retrograde stars shows the most significant decrease with time. Interactions of the binary with retrograde stars may, therefore, cause a more efficient decrease in eccentricity, circularizing the SMBHB orbit. This effect has already been observed by \cite{Sesana11}, who found that counterrotating stars are able to extract angular momentum more efficiently from the binary, leading to a faster decrease in its eccentricity.  While M4 and M5 show the smallest final eccentricities, the final separation of their SMBHBs is the largest among our models. The bottom panel of Fig. \ref{fig:fract_retrograde}  shows the inverse of the binary separation rescaled to its final value. Although counterintuitive, this representation effectively illustrates how in M4, the separation quickly decreases between 15Myr and 20Myr, when the merger is complete. In M5, the separation has a different behaviour; it reaches 30\% of its final value very quickly at the end of the merger, and then slowly decreases to its final value. In M1, the evolution of the separation shows a delay with respect to the end of the merger. While the merger is complete after around 2~Myr, the separation starts to decrease only after 7~Myr from the beginning of the simulation. The SMBHB in M1 has the lowest final separation ($1.8\times10^{-3}$~pc) among our models. M2 and M3 show a very similar final separation ($\sim 3.0\times10^{-3}$~pc) and similar linear decreases after the end of the merger. The separation seems to be correlated to the SMBHB mass ratio and the initial angular momentum, as concluded in \cite{Mastrobuono23}. A greater number of retrograde stars appears to be associated with reduced final separation values.  All the binaries are in a hard state at the end of the simulations, and they are expected to merge within a time frame of less than 5.7 billion years, as estimated by \cite{Ogiya20}.
Notably, we found a strong correlation between angular momentum and the eccentricity of the SMBH binary. The component of angular momentum perpendicular to the rotation plane and calculated with respect to the density centre of the system, $L_z$, decreases during the merger (see Figs B1-B5). In models M1–M5, it experiences a sharp drop when the two NSCs are finally destroyed and merged, going through a phase of violent energy and angular momentum redistribution. After this phase, the system stabilizes, and $L_z$ increases to reach its final value. As $L_z$ starts to rise after this drop, the binary forms, and the eccentricity reaches its final value.

\subsection{Shape and spatial distribution of the prograde and retrograde subpopulations}
 Figure \ref{fig:den_maps} shows the mass maps for the prograde and retrograde stars in each of our models, after 20~Myr of evolution. The mass maps for the entire NSC in analogue simulations can be found in \cite{Mastrobuono23}. The two kinematic groups of stars have remarkably different spatial distributions. While the prograde stellar component is distributed in a flattened ellipsoidal structure which spans over the entire size of the final NSC, the bulk of the retrograde component is less spatially extended and approximately spherical. The half-mass radii of the two components and the whole NSC are reported in Table \ref{tab:tab1}. As seen in \cite{Mastrobuono23}, the total half-mass radius increases with the SMBHB mass ratio and with the initial cluster distance (i.e. with the amount of orbital angular momentum). Among the models, M6 and M7, which either have no or only one SMBH, exhibit the smallest half-mass radii, while M5, the model with the highest $q$, shows the lowest central surface density. The retrograde population always shows a smaller half-mass radius ($r_{h,r}$) than the prograde population ($r_{h,p}$). The radius $r_{h,r}$ can be up to $2.5$ times smaller than  $r_{h,p}$. This happens in the case of M4, which is the model that starts with a larger initial distance between the two NSCs. Model M1 (with a small $q$) shares similarities with models M6 and M7. In all of these cases, the value of $r_{h,r}$ is approximately half that of $r_{h,p}$.
 Model M2 exhibits the closest values for the half-mass radii of the two subpopulations, likely due to the limited availability of angular momentum in the progenitor NSCs. In M3, $r_{h,r}$ is again half of $r_{h,p}$. In M5 (with a large $q$), $r_{h,r}$ constitutes about 60\% of $r_{h,p}$, with both values being the largest among all models. This suggests that a significant SMBHB mass ratio results in the largest systems, even when considering different kinematic groups. 

The density profiles of the two populations show differences that depend on the initial conditions of the model. Figure \ref{fig:density} shows the density profiles of the two populations and of the entire NSC. The bullets show the final distance from the centre of the two (or single) SMBHs, when present. The distance of the closest retrograde and prograde stars to the centre of the final NSC are listed in Table \ref{tab:tab2} for each of the models with two SMBHs. We note here that when we write `star' in reality we mean a particle representing a mass of $152\,M_{\odot}$, so that they convey a statistical picture of the evolution, of the orbits. We cannot tell what particular orbit a given star is going to follow, since the results we are presenting are to be envisaged statistically.
In M1, the closest stars to the SMBHB are on prograde orbits, and reach a distance of 0.017~pc from the NSC centre. The prograde stars show a central cuspy rise in the density profile, due to the presence of three stars in the first two bins. The same trend is confirmed when using a smaller number of bins for building the density profile. Again, talking about a cusp with three stars would be ill-defined, but in total, we have in this region about $500\,M_{\odot}$ in stellar matter. At this kind of radius, the density profiles are dominated by main-sequence stars which, on average, have, at most, one solar mass each, so we deem it reasonable to interpret the cusp as such. The density profile of the retrograde stars overlaps with that of the retrograde stars between 0.03~pc (the radius where the prograde stars exhibit the cusp upturn)  and 0.3~pc. Both profiles have a slope of about $-1.2$ in this radial range. At larger radii, the prograde stars show larger densities and a smaller slope than the retrograde stars.

Reducing the initial orbital angular momentum significantly alters the density profile of both components. In model M2, the nearest star to the SMBHB is retrograde and is found at a distance of $3.5\times10^{-3}$~pc. Therefore, the cuspy upturn of the density profile due to the presence of this star is only observed in the retrograde component. The two density profiles overlap up to 0.4~pc and have a slope of $-0.9$. At larger radii, prograde stars are more numerous and dense than retrograde stars.
In model M3, the closest stars to the binary are on prograde orbits, with the closest star found at $1.8\times10^{-2}$~pc from the NSC centre. Both the prograde and retrograde stars show a central density cusp with slope $\sim -1.5$, followed by a core-like profile. The prograde stars start to dominate at radii larger than 0.5~pc. 

In M4, stars on prograde orbits reach closer distances to the centre of the NSC than retrograde stars. Both retrograde and prograde stars can reach distances that overlap with the SMBHB orbit. The density profile shows a central cuspy uprise for both subpopulations. This feature is due to the presence of one prograde star in the first density bin and two retrograde stars in the second and third bins. Between 0.05~pc and 0.4~pc the prograde stars show a cuspy density profile (with slope $\sim -1.35$), and the retrograde stars have a flat density profile. While the prograde stars are denser at any radius, at radii larger than 0.4~pc the prograde stars significantly dominate over the retrograde stars. The closest stellar particle to the binary is at $4.3\times10^{-3}$~pc from the centre. 

In M5, the closest stars to the binary are found at larger radii with respect to the other cases. The closest star to the binary is at only $0.11$~pc from the centre, and it is in a prograde orbit. Both the prograde and retrograde stars show a core-like density profile. This model shows the lowest central density value. 

Similarly, M6, which has no SMBH, shows cored profiles for both subpopulations. However, the density value is higher in M6 than in M5 and the two subpopulations show the same central value of density, while in M5 the prograde stars have higher central density at any radius compared to the retrograde stars.

In M7, prograde stars arrive extremely close to the single SMBH, with a minimum distance of $1.1\times10^{-3}$~pc. The central cuspy upturn observed in the prograde component is due to the presence of the closest star to the centre. Both subpopulations show a central cusp (with slope $\sim -1.35$).  The two profiles are comparable up to 0.6~pc. At larger radii, the prograde component is denser than the retrograde one.
In all cases, the density profile of the two populations evolves with time. Remarkably, in the case of M1, a steep cusp is present in both components at 16~Myr. This cusp is later destroyed in the interaction by the SMBHB, which at the end of the simulation becomes tight enough to eject stars from the central regions of the system. In M5, which shows the flatter density profiles among the models with two SMBHs, the SMBHB very quickly becomes tightly bound; a cusp forms in the two profiles to be then scoured by the action of the SMBHB. Between 4 and 16~Myr, a slight decrease in the density is observed at the centre of the two profiles. This decrease is followed by the rise of the cusp at radii smaller than 0.1~pc. 

\subsection{Axial ratios}
We estimated the shape of the two populations using their axial ratios. To do so, we used the iterative method introduced by \cite{Katz91} and calculated the principal components of the inertia tensor. In this way we identify the symmetry axes of the particles inside the spheroid of radius $r^2 = x^2/a^2 + y^2/b^2 + z^2/c^2$, setting a precision of $5\times10^{-4}$, with $a$, $b$  and $c$ as the major, intermediate and minor axis of the ellipsoid, respectively.

The results of this analysis are presented in Fig. \ref{fig:ax_ratios}. 
The prograde component is triaxial in the central 5~pc when two SMBHs are present. The triaxiality increases with $q$. When only one or no SMBHs are present, the central regions of the NSC are spherical, and so is the prograde component. 
Outside the central few parsecs, the prograde stars are oblate, with a flattening between 0.5 and 0.7, depending on the initial conditions of the merger, as seen in \cite{Mastrobuono23} for the entire NSC. The retrograde component is approximately spherical outside the central 2-5~pc. The deviation from sphericity depends on the merger conditions and also on the distance from the cluster centre. A large $q$ leads to a more spherical retrograde component. M4 and M7 show the largest deviation from sphericity. When $q\leq 0.1$ or when only one SMBH is present, the system becomes more spherical at larger distances from the centre. In the other cases, $c/a$ is either almost constant or increases with the radius. 

\subsection{Kinematics of the prograde and retrograde subpopulations}
We analysed the kinematic properties of the prograde and retrograde stellar components. Figure \ref{fig:rot_curves} shows the rotation curves of the two kinematic components and of the entire NSC for each of the models, at the end of the simulations. The retrograde stars always rotate slower than the prograde stars. The peak velocity of the prograde population is always around 40~km/s, and it is reached between 0 and 5~pc from the centre of the system. This peak velocity is slightly lower in M2 and more significant in M7. The peak velocity and the shape of the rotation curve exhibit significant variability within the retrograde population. This peak velocity consistently falls within the range of 20-30~km/s and is reached closer to the centre of the NSC compared to the prograde stars. The lowest value is observed in the scenario where no SMBHs are present in the NSCs (M6). While, in some scenarios, the rotational velocity of retrograde stars significantly decreases as one moves radially outwards, in the case of M2, their rotational velocity remains close to the peak velocity up to a distance of 20~pc from the centre of the NSC. This, in conjunction with the lower rotational velocity of prograde stars, results in a system characterized by a low total rotational velocity, consistent with the expectations for a system with low initial orbital angular momentum. Except for the case with only one SMBH (M7), the curves exhibit approximate symmetry with respect to the cluster centre.
In M7, both the prograde and retrograde stars reach their peak velocity very close to the centre of the NSC, and the retrograde stars curve shows a significant amount of noise in the external regions. 
All these figures are obtained considering the NSC as seen edge-on, that is, with the line of sight perpendicular to the rotation axis. The rotation curves are obtained by putting a mock slit at the cluster centre, perpendicularly to the rotation axis, with an extension along $z$ of 2~pc. 

Figure \ref{fig:dispersion_curves} shows the velocity dispersion of the two components in each of the simulated models. The velocity dispersion shows a high degree of variability compared to the rotation curves. In M1 the central velocity dispersion for the two components has a difference of about 10~km/s and the total velocity dispersion is dominated by the retrograde stars. While the central velocity dispersion is higher for the retrograde stars, outside 5~pc the situation is reversed, with the prograde stars showing a higher velocity dispersion compared to the retrograde stars. In M2, the retrograde stars' central velocity dispersion is more than 40~km/s higher than for the prograde stars. The two velocity dispersions are comparable outside 4~pc. The entire NSC has a central velocity dispersion slightly higher than for the prograde stars. In M3, the prograde and retrograde stars have comparable velocity dispersions throughout the system. In M4, the prograde stars have a central velocity dispersion which is much higher (200~km/s) than the retrograde stars (90~km/s). Outside the central 5~pc the two velocity dispersions are comparable. The entire system has a high central velocity dispersion, comparable to that of the prograde stars. Among the systems with two SMBHs, M5 shows the lowest values for the velocity dispersion of the prograde and retrograde stars separately. The velocity dispersions of the two components are comparable, with the total velocity dispersion being higher than that of both components, outside the central bin. In M6, the absence of any SMBHs results in a broader shape of the velocity dispersion curve, with the central value slightly higher for retrograde stars compared to prograde stars. The central peak of the velocity dispersion for the entire system is the widest among all simulated systems.
M7 shows a very peculiar velocity dispersion curve, with a sharp rise at the centre and values below 50~km/s outside the central few parsecs. The rise is extreme ($\sim1500$~km/s) for the prograde component and the entire system ($\sim1150$~km/s), compared to the retrograde component ($\sim100$~km/s). This extreme rise in velocity dispersion is due to the scattering action of the single SMBH present at the centre of the system, and the values observed are comparable to the velocity dispersion detected for the S-stars at the Galactic centre \citep{Genzel10}. Retrograde stars seem to have milder interactions with the central SMBH, which leaves them with a lower dispersion.
Additional information on the kinematic state of the NSC is provided by the velocity dispersion and $V_{LOS}/\sigma_{LOS}$ parameter, which quantifies the rotational support of a rotating system. Figure \ref{fig:Vsigma} shows the $V_{LOS}/\sigma_{LOS}$ for all our models. All systems are rotationally supported, with M2 showing the lowest total $V_{LOS}/\sigma_{LOS}$ value, due to the lowest initial amount of orbital angular momentum that leads to the production of more retrograde stars. The other curves do not show any significant dependence on the kind of central object present in the system. The total $|V_{LOS}/\sigma_{LOS}|$ is, in all cases, about equal to unity at radii larger than a few parsecs, and the prograde component is always more rotationally supported than the retrograde component.

In addition, we calculated the velocity anisotropy parameter for the different components and for the entire system. The $\beta$ parameter is defined as
\begin{equation}
    \beta(r)=1-\frac{\sigma_\theta(r)^2+\sigma_\phi(r)^2}{2\sigma_r(r)^2}
,\end{equation}
where $\sigma_r$, $\sigma_\theta$, and $\sigma_\phi$ are the components of the velocity dispersion in spherical
coordinates. If the system is isotropic, $\beta$ is equal to zero. If the kinematic of the system is dominated by radial orbits $\beta>0$, while the majority of the stars are on tangential orbits $\beta<0$. In the limit of all circular orbits, $\beta=-\infty$. 

The results of this analysis are presented in Fig. \ref{fig:beta_curves}. 
As found in \cite{Mastrobuono23}, when two SMBHs are present, the system is tangentially anisotropic at the centre and becomes radially anisotropic going at larger distances from the centre. When only one or no SMBHs are present, the system is isotropic at the centre and becomes radially anisotropic going outwards. In all cases, the entire system is more isotropic than the two components taken separately. This occurs because, individually, prograde and retrograde stars exhibit lower tangential velocity dispersion compared to the radial dispersion. However, when considered together, the tangential velocity dispersion increases due to the differing rotational velocities of the two stellar components, amplifying their tangential dispersions. Except for model M2, in which the two components have similar behaviours, the prograde stars are more isotropic than the retrograde stars. The behaviour of the anisotropy of the two components at the NSC centre varies significantly with the merger parameters. Generally, the anisotropy of the retrograde component is radial at the centre, then it moves towards isotropy to eventually become radially anisotropic. Prograde stars can be radially (M3 and M7) or tangentially anisotropic (M1, M2, M5, and M6) or even isotropic (M4), depending on the merger characteristics. {In addition, the mean eccentricity of the retrograde stars, which is between $0.71$ and $0.73$, depending on the cases, is slightly closer to unity than for the prograde stars which range between $0.65$ and $0.68$. Therefore, we do not observe large differences in the orbital circularity of the two populations.}

Finally, Figs. C1 and C2 show the velocity maps of all systems, as would be obtained through IFU observations. The figures are obtained by applying the Voronoi binning procedure described by \cite{Cappellari03} aiming at a signal-to-noise ratio (S/N) of $15$ in each
bin. From the maps, it is apparent that the rotation of the NSC is the result of the superposition of the rotation of the two populations. In all models, the prograde stars are rotating with a similar pattern, while the retrograde component is more or less extended, as described before, and rotates at different speeds depending on the merger conditions. Most of the difference seems to arise from the amount of orbital momentum, with a second-order contribution from the SMBH mass ratio. The high-velocity dispersion due to the presence of the high mass-ratio binary or of the single SMBH is clearly visible in these plots as high-velocity holes carved by those systems in the centre of the map. While the velocity and dispersion curves can be compared with observations done with slits, the maps can be compared to IFU observations, such as those that can be obtained with MUSE.



\section{Conclusions}

Nuclear cluster mergers, involving the fusion of dense stellar clusters near the centres of galaxies, are essential in the formation and evolution of galactic structures. The distribution of stellar orbits within these mergers is vital for understanding underlying mechanisms and their broader astrophysical implications. 
In this study, we examine the orbital distribution of stars in merging NSCs and their impact on the evolution of SMBHBs. We employed various numerical models to investigate different dynamical aspects.

Initially, at time $t=0$, all our models do not have any initial internal rotation but display varying amounts of orbital angular momentum since they are orbiting around each other on different relative orbits. We considered various initial conditions, such as the SMBH mass ratio, initial eccentricity, and the presence or absence of a central SMBH. These conditions play a decisive role in determining the final distribution of stars and the subsequent evolution of the SMBHB. We followed all mergers up to 20~Myr, as computational limitations arise after the hard binary's formation, preventing us from continuing the run. 
When the binary becomes hard, the timestep becomes increasingly small, leading to an extremely long computational time. According to an analytic estimate, the SMBH coalescence time mostly depends on the SMBH mass ratio and varies between 57.6~Myr for $q=0.01$ and 5.3~Gyr for $q=1.0$ \citep{Ogiya20}. However, we expect this coalescence, as well as the two-body relaxation effects,  to mostly affect the very central regions of the system, which are not yet observationally accessible, leaving the regions outside the few central parsecs almost unchanged. In all cases, the relaxation time at the radius of the SMBH sphere of influence (2-3~pc), or at the half-mass radius of the system when the SMBH is not present\footnote{The half-mass relaxation time was estimated using the formula taken from \cite{Spitzer87}.}, is larger than the Hubble time \citep{Merritt2010, Antonini12, Tsatsi17, Abbate18}. Therefore, the snapshot taken at 20~Myr can be considered as representative of the long-term structural properties of the system, at least outside the few central parsec.

The final structural and kinematic properties of the simulated systems strongly depend on the merger parameters, as already observed in \cite{Mastrobuono23}. Here, we focus on the presence of retrograde and prograde stars and their dynamical signatures. All models start with a similar fraction of retrograde stars, depending on the internal and orbital angular momentum of the merging systems. We observe that stochastic gravitational encounters between stars randomly invert stellar angular momenta, most likely transforming a retrograde star into a prograde one.  During the interaction between the two NSCs, stars exchange energy and angular momentum, driving the system towards equilibrium. This results in part of the initially retrograde stars aligning their spin with the overall cluster rotation (see Figs B1-B6). Simulations with no, one, and two SMBHs show similar behaviours, suggesting that the binary has a minimal role in the evolution of the retrograde stellar fraction. In addition, the two SMBHs, when present, become bound after the fraction has approximately reached its final value (see magenta cross in Figs B1-B5).  At the end of the simulations, model M5, which has the largest SMBH mass ratio, shows the lowest fraction of retrograde stars, while model M2, which has the lowest initial orbital angular momentum, shows the largest fraction of retrograde stars. Therefore, the mass ratio and orbital conditions significantly impact the final fraction of prograde and retrograde stars. Prograde stars come in equal measure from the two progenitor NSCs, while the fraction of retrograde stars coming from the two progenitors can vary significantly. In most of the models, the majority of retrograde stars were in the primary NSC at the beginning of the simulation. Retrograde stars coming from the primary NSC are, therefore, more likely to maintain their initial angular momentum orientation when they arrive at the centre of the newly formed NSC.

We find that models with a larger variation of the retrograde star fraction show a more efficient decrease in eccentricity, resulting in SMBHBs with more circular orbits \citep[and see also][]{Sesana11}.  Interestingly, the binary's eccentricity is strongly correlated with the component of angular momentum perpendicular to the rotation plane ($L_z$). The eccentricity decreases and stabilizes as $L_z$ reaches its minimum value, a result of the forces and torques acting during the merger. Once the system fully merges, $L_z$ sharply increases, while the eccentricity reaches its final value \citep[see also][]{Tsatsi2015}. 

As found in \cite{Mastrobuono23}, the final SMBHB separation is linked to the SMBHB mass ratio and the initial angular momentum. Additionally, larger fractions of retrograde stars seem to correlate with smaller final separations. In all our final models, the distribution of retrograde stars is less extended and has a spherical spatial distribution, while prograde stars are distributed in a more extended disc. The half-mass radius of the prograde stars is up to 2.5 times larger than that of the retrograde stars, which are, therefore, less extended in radius than the prograde stars. Except for M2, which is characterized by a low initial orbital angular momentum, the closest star to the centre is on a prograde orbit. 

The shape of the density profile differs for the prograde and retrograde stars. The difference is more pronounced in the very central regions, where we observe cuspy upturns in one or both of the components, depending on the initial conditions, and at radii typically larger than 0.5~pc, where the prograde stars dominate over the retrograde stars. The shape of the two populations is quantitatively confirmed by their axial ratios. The triaxiality of the final system correlates with $q$, while when only one or no SMBHs are present, the central system is spherical. At radii larger than a few parsecs, the distribution of the prograde stars is oblate, and that of the retrograde stars is approximately spherical, with the sphericity increasing with $q$, in the case of two SMBHs.

The systems and their components also differ from a kinematic point of view. The NSCs that form from the merger rotate differently depending on the initial merger properties. The retrograde component always rotates slower than the prograde one. While the peak velocity for the prograde component is always around 40~km/s and is reached at 2-5 pc from the NSC centre, the peak velocity for the retrograde component is lower and shows more significant variability (20-30~km/s) and is reached at smaller distances from the centre. The line-of-sight velocity dispersions show significant variability between the different systems. Depending on the initial conditions of the merger, the velocity dispersion can be centrally dominated by the retrograde or prograde stars. Outside the central regions, the prograde stars generally exhibit a larger dispersion than the retrograde stars. When a binary is present and $q$ is smaller or equal to $0.1$, the central peak of the velocity dispersion is narrow (less than $5$~km/s). The peak widens for $q=1.0$ and when only one SMBH is present. When no SMBH is present, the prograde stars have a much higher velocity dispersion compared to that of the retrograde stars, and the velocity dispersion peak is extremely narrow. The $V_{LOS}/\sigma_{LOS}$ parameter shows that all the systems are strongly rotating, with M2 showing the lowest total $V_{LOS}/\sigma_{LOS}$ value, a fact linked to the lower initial orbital angular momentum and, possibly, to the smaller number of retrograde stars produced in this run. No significant dependence on the presence or absence of a central SMBHB is observed. The prograde component is always more significantly rotating  than the retrograde component. When two SMBHs are present, the system is centrally tangentially anisotropic and becomes radially anisotropic going outwards. The presence of only one or no SMBH is related to more centrally isotropic systems. When taken separately, the prograde and retrograde components are more anisotropic than the entire system, due to their individual tangential velocity dispersion, which is lower compared to the total one. The prograde stars are generally more isotropic than the retrograde stars, and the radial behaviour of the two components varies significantly with the merger parameters.

All the dynamical and structural properties of the two kinematic populations described in this work, as well as the fraction of prograde and retrograde stars and their orbital parameters that can be found in real NSCs, can be ultimately used to `reverse-engineer' the observations to indirectly detect the presence of an SMBHB and to reconstruct the properties of the two progenitor NSCs, including their chemical properties, as well as the orbital conditions of the merger, the SMBH mass ratio, and the properties of the merging galaxies. 

Finally, there is another particularly interesting application of our study. The orientation of these orbits, whether prograde or retrograde, has significant effects, in particular, on the formation and characteristics of EMRIs, as put forward by the work of \cite{Amaro-SeoaneEtAl2013}, as the location of the ISCO depends both on the SMBH spin and on the orientation of the stellar orbit relative to the black hole's spin axis \citep{Amaro-Seoane2013}. This has practical applications, such as predicting and interpreting GW signals, and has implications on the event rate of EMRIs, which is important for future GW observatories such as LISA \citep{Amaro-SeoaneEtAl2017}. However, the difference of velocity dispersion has  yet to be addressed in order to interpret, or rather translate, these results into event rates. Once the SMBHB enters the GW dominate regime, the evolution timescale is determined by the characteristic
gravitational radiation timescale, proportional to the semi-major axis to the power of four. This leads to a prompt coalescence from the standpoint of the stellar population. In a timescale that is orders of magnitude smaller than the relaxation or even dynamical times, the two SMBHs merge and become one. How the stellar system adapts to this change within the sphere of influence and the impact on the stellar orbits as well as the architecture of the NSC will be addressed elsewhere.


\section*{Data availability}
Appendices B and C and the relative figures cited in the text are available at: \url{https://zenodo.org/records/13986686}

\begin{acknowledgements}
 We thank the referee for their thoughtful and constructive feedback, which has greatly improved the quality of this work. 
 AMB acknowledges funding from the European Union’s Horizon 2020 research and innovation programme under the Marie Sk\l{}odowska-Curie grant agreement No 895174.
 This research is funded by the Science Committee of the Ministry of Education and Science of the Republic of Kazakhstan (Grant No. AP23487846). This research has been funded by the Aerospace Committee of the Ministry of Digital Development, Innovations and Aerospace Industry of the Republic of Kazakhstan (Grant No. BR20381077). This research has been/was/is funded by the Committee of Science of the Ministry of Science and Higher Education of the Republic of Kazakhstan (Grant No. BR24992759 ).
 MJFA has been supported by the Spanish Ministerio de 
Ciencia, Innovación y Universidades and the Fondo Europeo de 
Desarrollo Regional, Projects PID2019-109753GB-C21 and 
PID2019-109753GB-C22, the Generalitat Valenciana, grant CIAICO/2022/252
and the Universitat de València Special Action Project 
UV-INVAE19-1197312.
 GO was supported by the National Key Research and Development Program of China (No. 2022YFA1602903) and the Fundamental Research Fund for Chinese Central Universities (Grant No. NZ2020021, No. 226-2022-00216).
 This work has made use of the computational resources obtained through the DARI grant A0120410154. All the models presented in this work have been run on the GPU-equipped supercomputer Jean-Zay at IDRIS, the national computing centre for the CNRS (``Centre national de la recherche scientifique''). PAS acknowledges the funds from the ``European Union NextGenerationEU/PRTR'', Programa de Planes Complementarios I+D+I (ref. ASFAE/2022/014).
 
\end{acknowledgements}

%
%
   \bibliographystyle{aa} 
   \bibliography{pro} 

\begin{appendix}
\section{Additional figures}
\begin{figure*}
   \raggedright
   \includegraphics[width=0.23\textwidth]{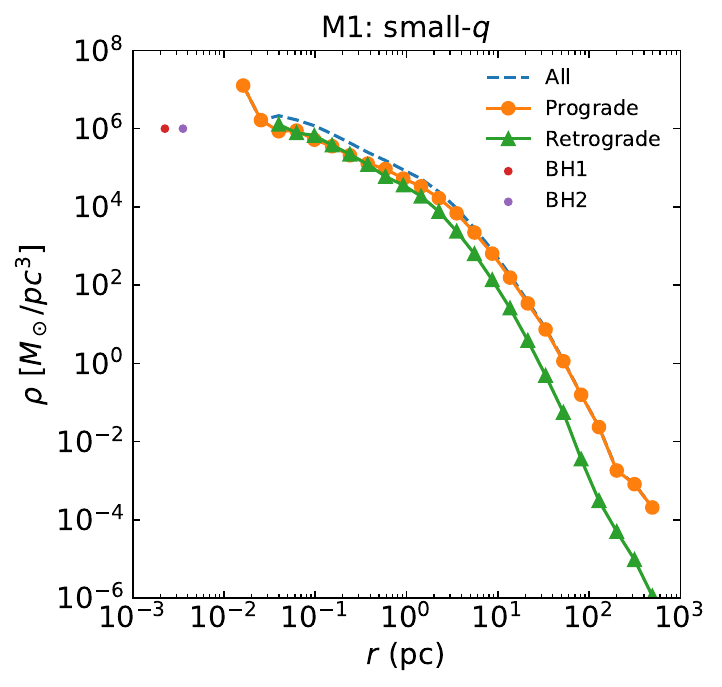}
   \includegraphics[width=0.23\textwidth]{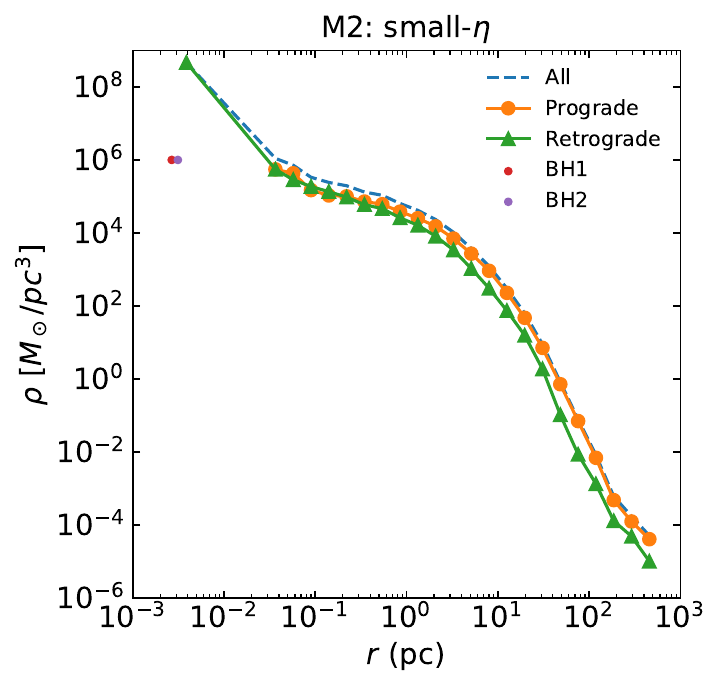}
   \includegraphics[width=0.23\textwidth]{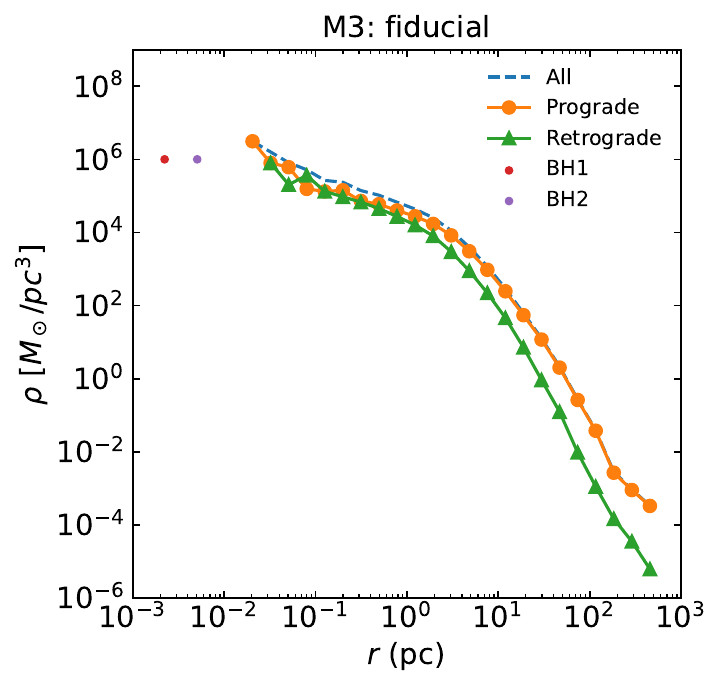}
   \includegraphics[width=0.23\textwidth]{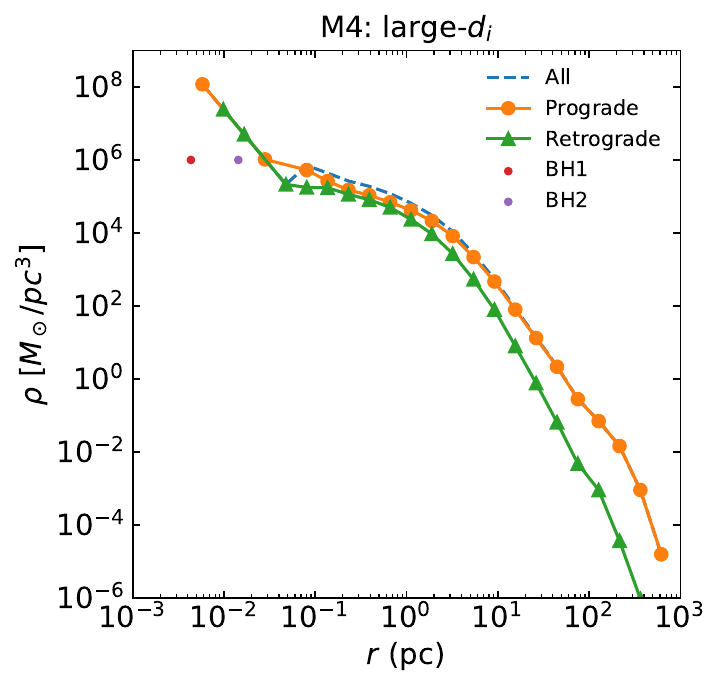}
   \includegraphics[width=0.23\textwidth]{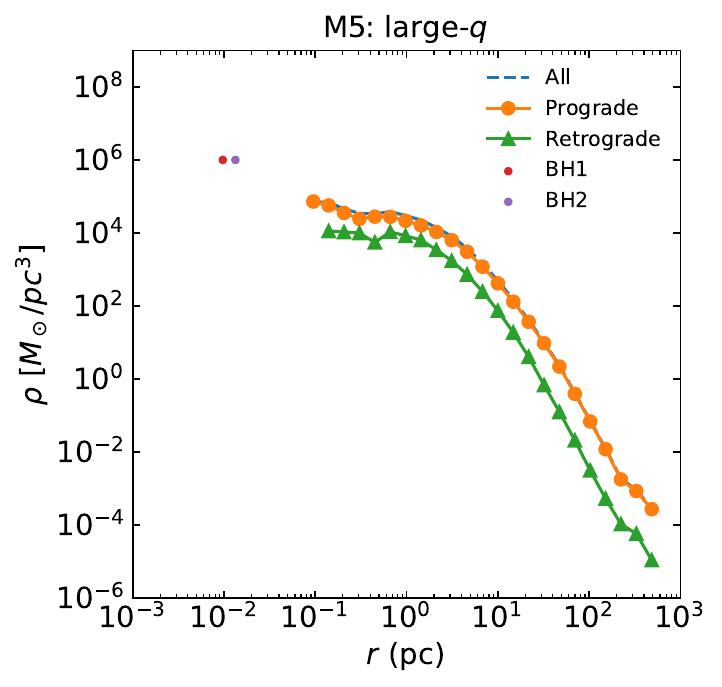}
      \includegraphics[width=0.23\textwidth]{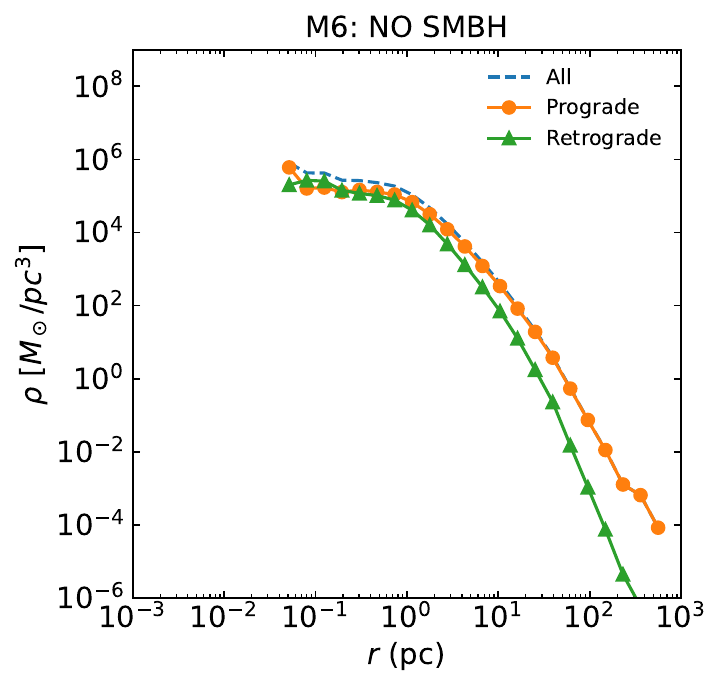}
         \includegraphics[width=0.22\textwidth]{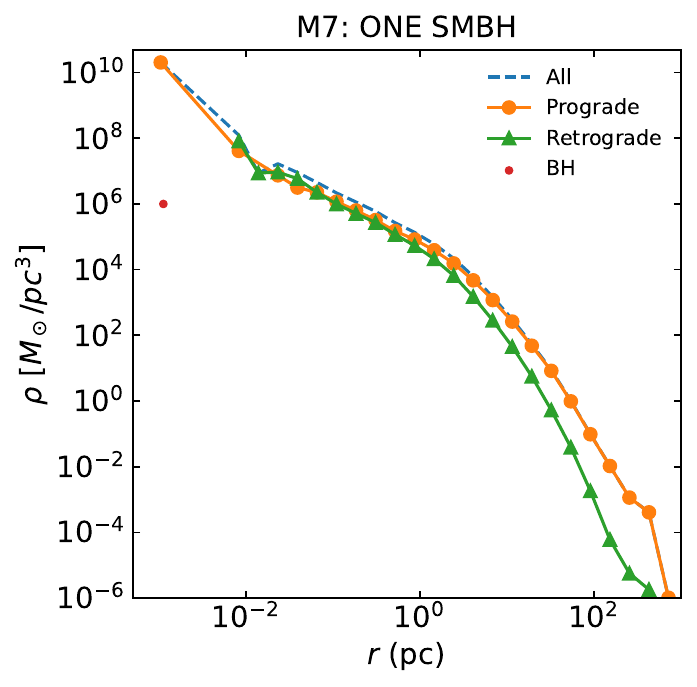}
      \caption{Spatial density of the prograde (orange line with bullets) and retrograde (green line with triangles) stellar components of the final NSC. The density profile of the whole NSC is also shown (blue dashed line). The final distance from the centre of the two SMBHs forming each binary, or of the single SMBH in the case of M7, is shown using one or two bullets. 
              }
         \label{fig:density}
   \end{figure*}   

\begin{figure*}
   \raggedright
   \includegraphics[width=0.23\textwidth]{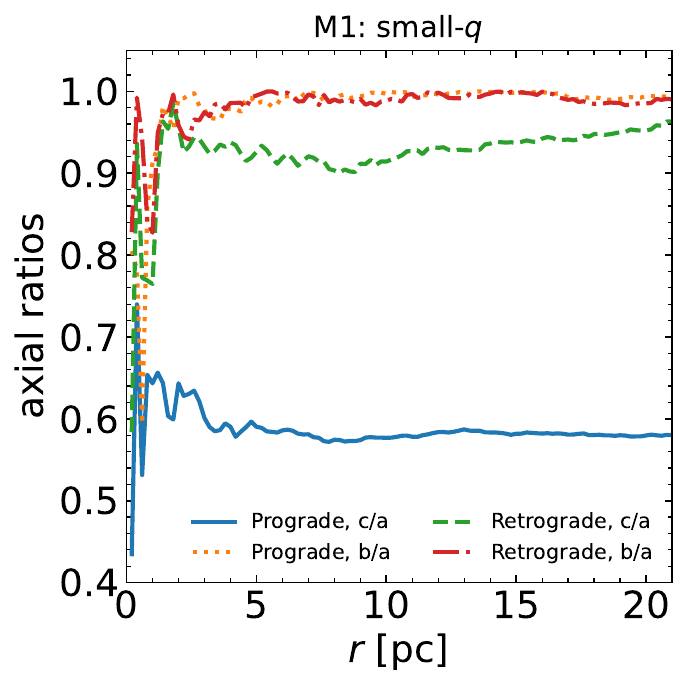}
    \includegraphics[width=0.23\textwidth]{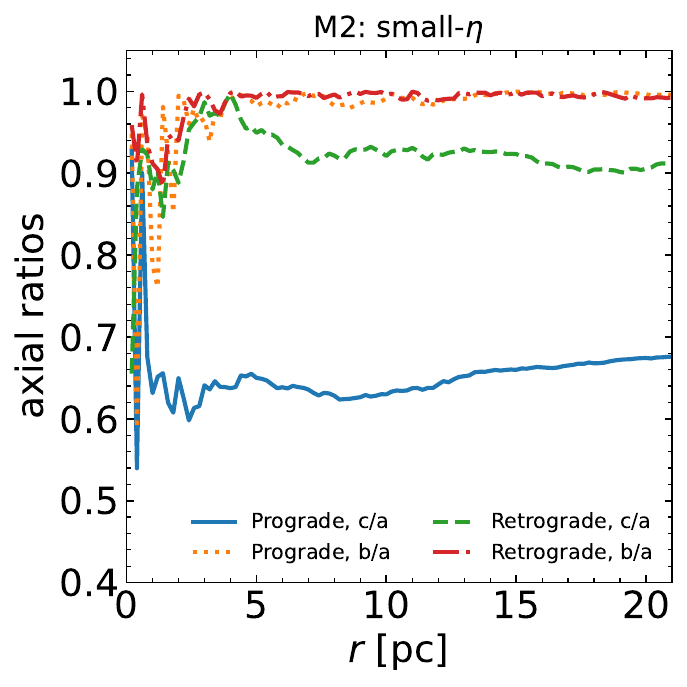}
    \includegraphics[width=0.23\textwidth]{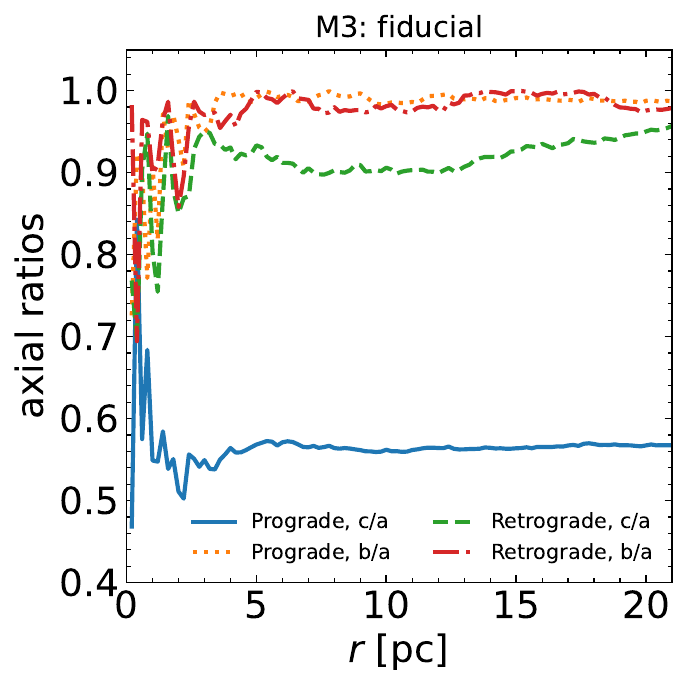}
    \includegraphics[width=0.23\textwidth]{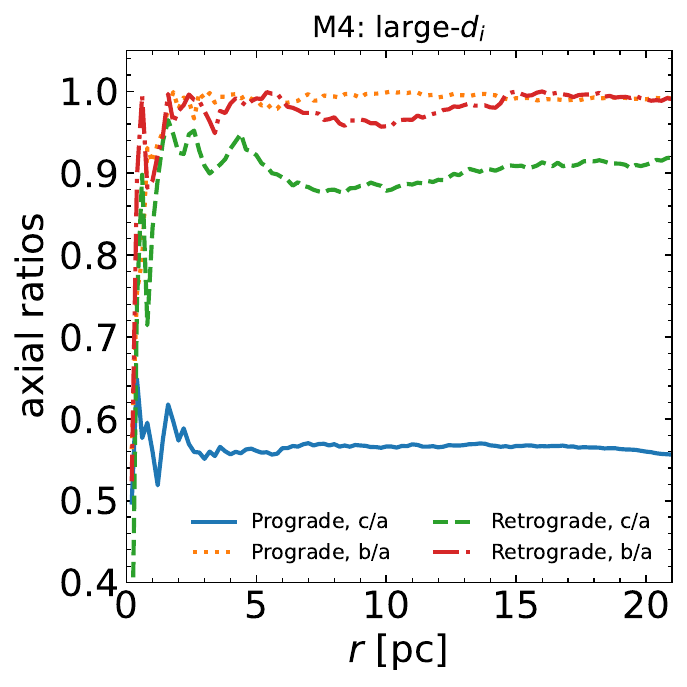}
    \includegraphics[width=0.23\textwidth]{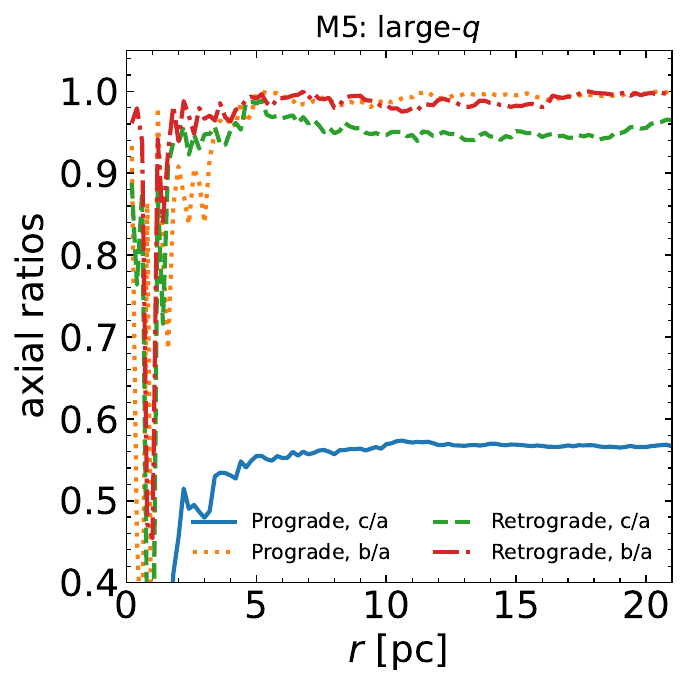}
    \includegraphics[width=0.23\textwidth]{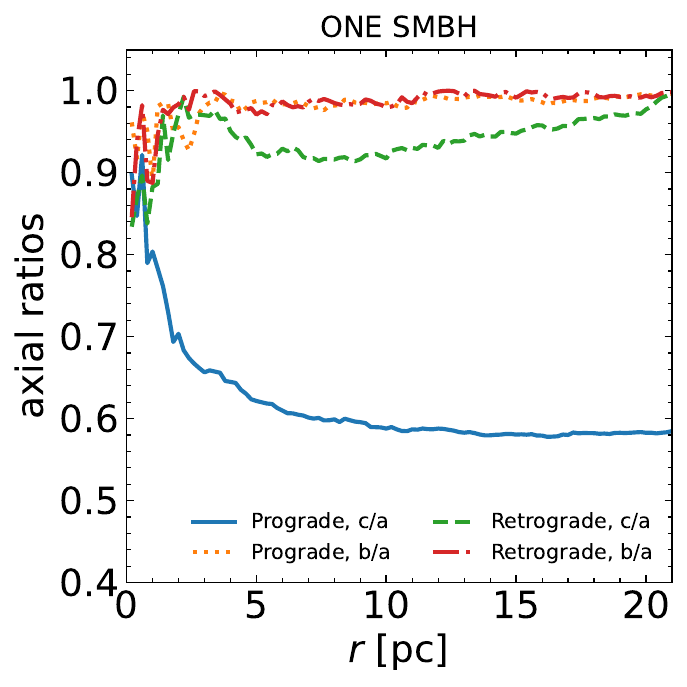}
    \includegraphics[width=0.23\textwidth]{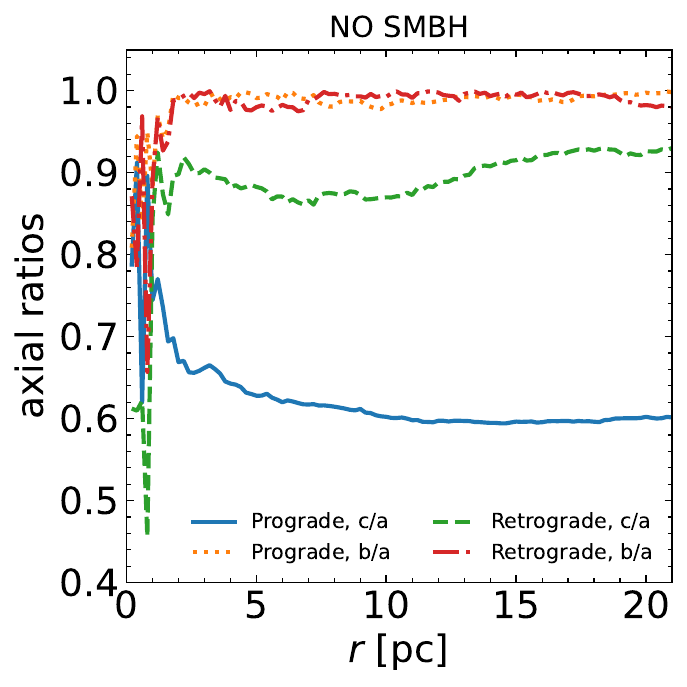}
      \caption{Intermediate (dotted orange line) minor (solid blue line) axial ratios of the prograde population. The same ratios are shown for the retrograde population (green dashed line and red dot-dashed line, respectively). The prograde population is oblate and significantly flattened. The retrograde population is approximately spherical while being mildly oblate.
              }
         \label{fig:ax_ratios}
   \end{figure*}

\begin{figure*}
\raggedright
    \includegraphics[width=0.22\textwidth]{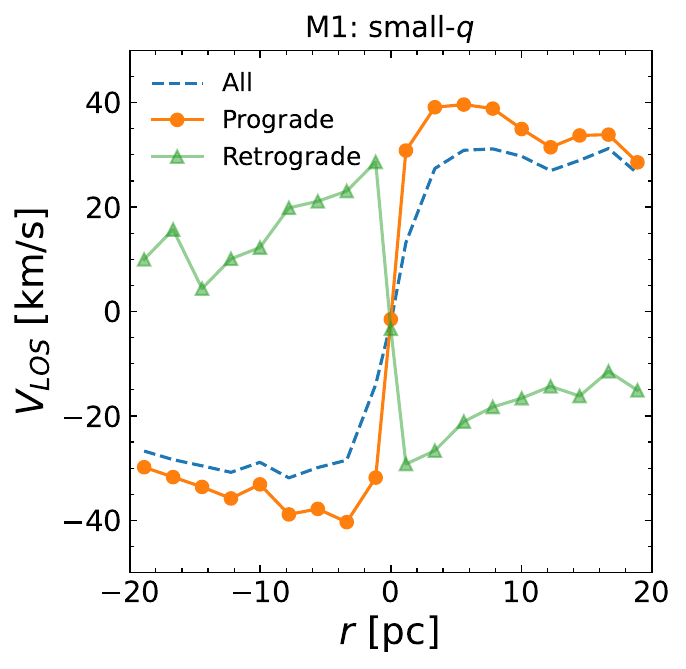}
    \includegraphics[width=0.22\textwidth]{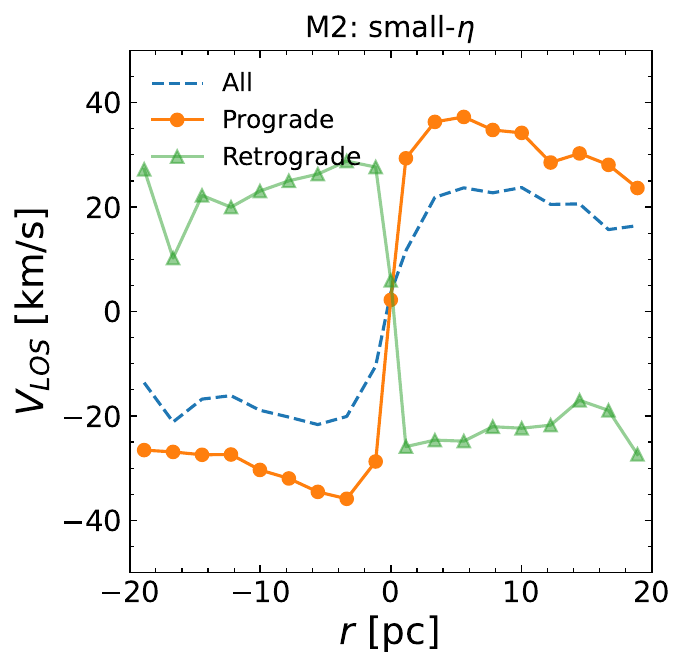}
    \includegraphics[width=0.22\textwidth]{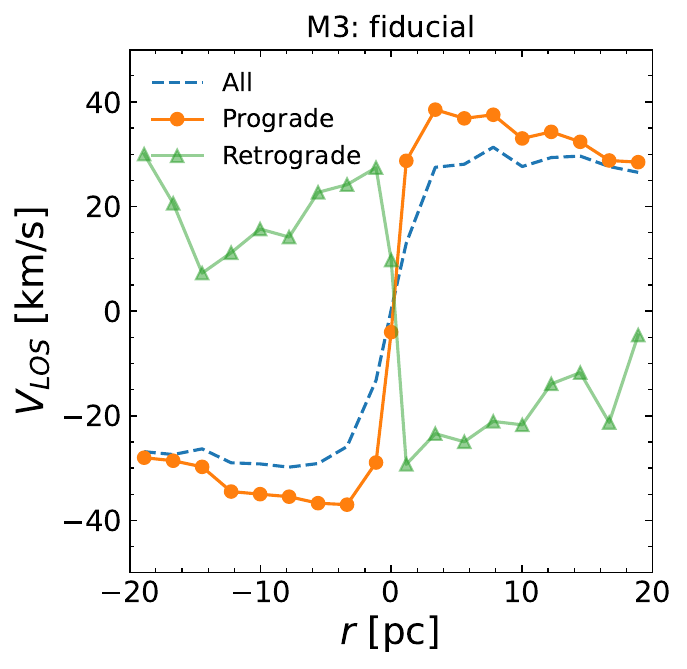}
    \includegraphics[width=0.22\textwidth]{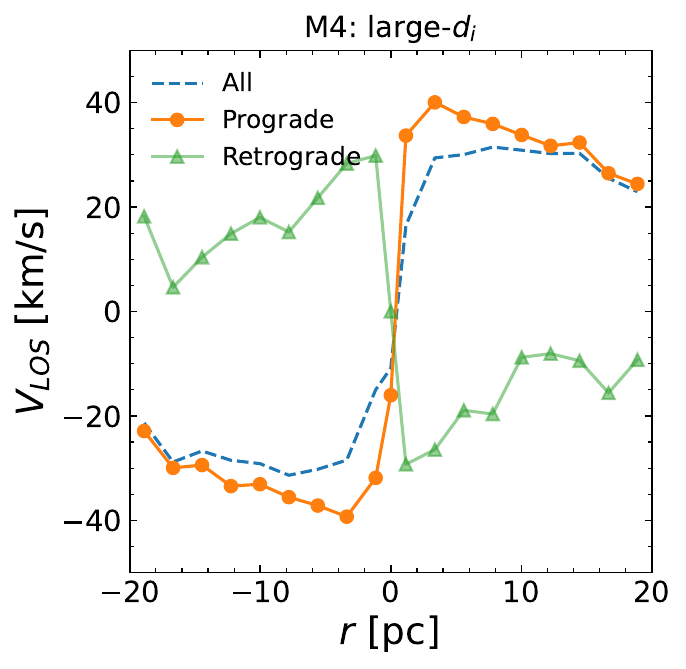}
    \includegraphics[width=0.22\textwidth]{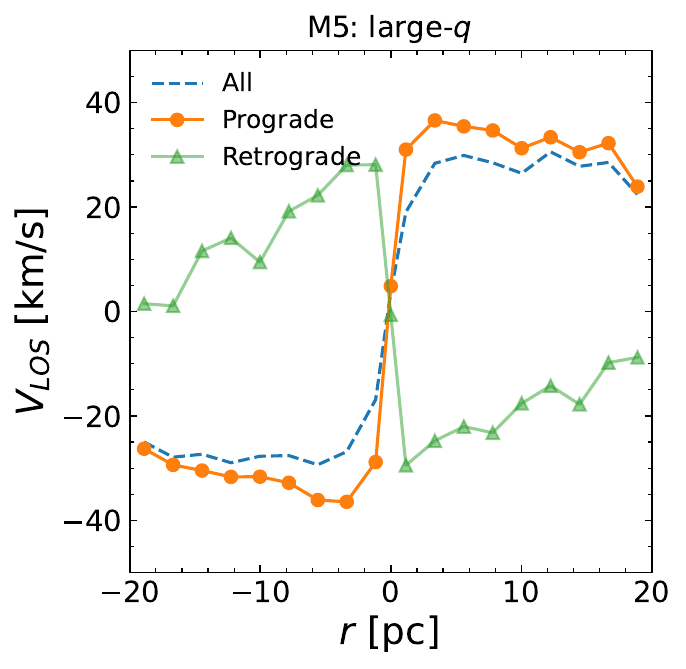}
    \includegraphics[width=0.22\textwidth]{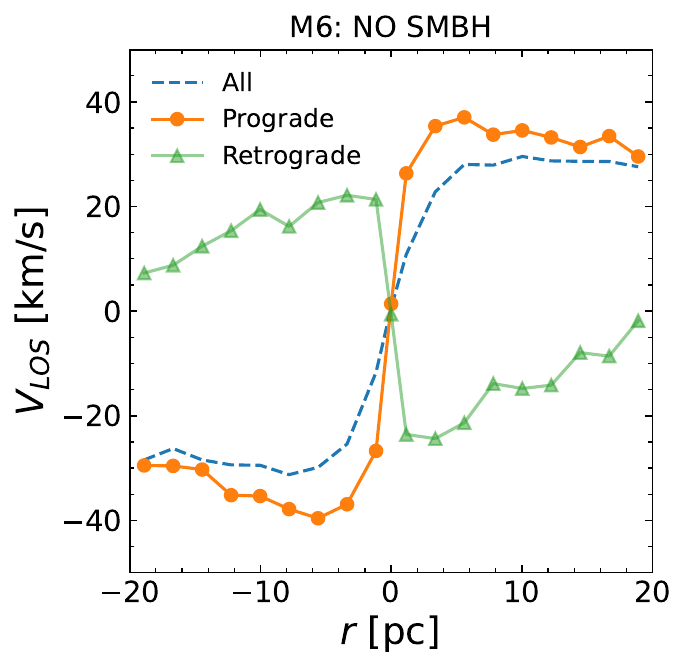}
    \includegraphics[width=0.22\textwidth]{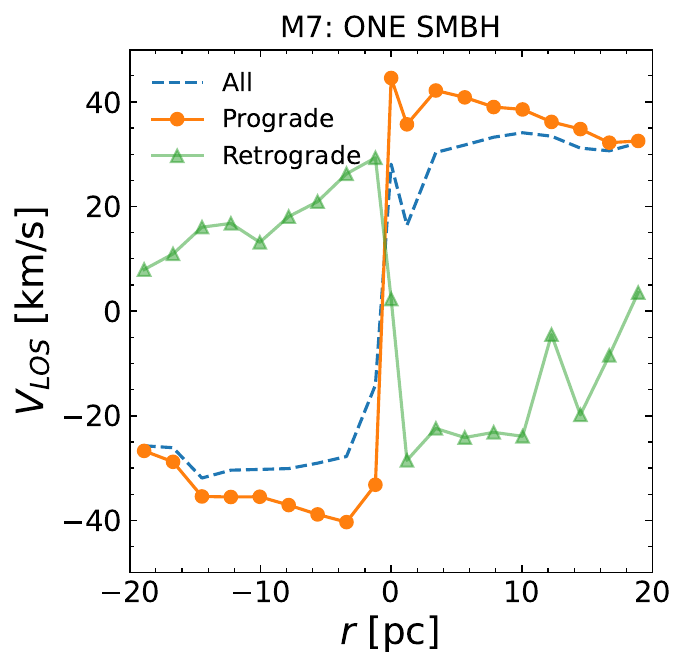}

    \caption{Velocity curves for the entire NSC (blue dashed line), of the prograde population (orange line with bullets), and of the retrograde population (green line with triangles). The prograde population always rotates faster and reaches the velocity peak at larger distances with respect to the retrograde one.}
    \label{fig:rot_curves}
\end{figure*}
\begin{figure*}
\raggedright
    \includegraphics[width=0.22\textwidth]{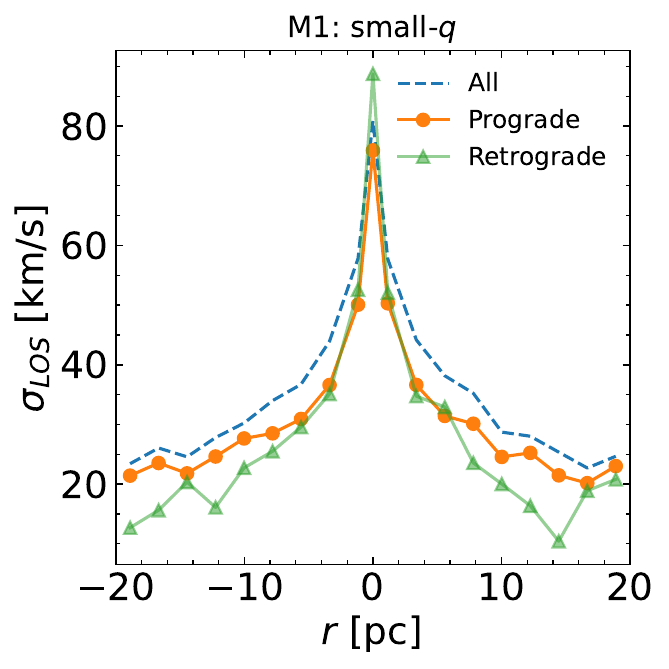}   
    \includegraphics[width=0.22\textwidth]{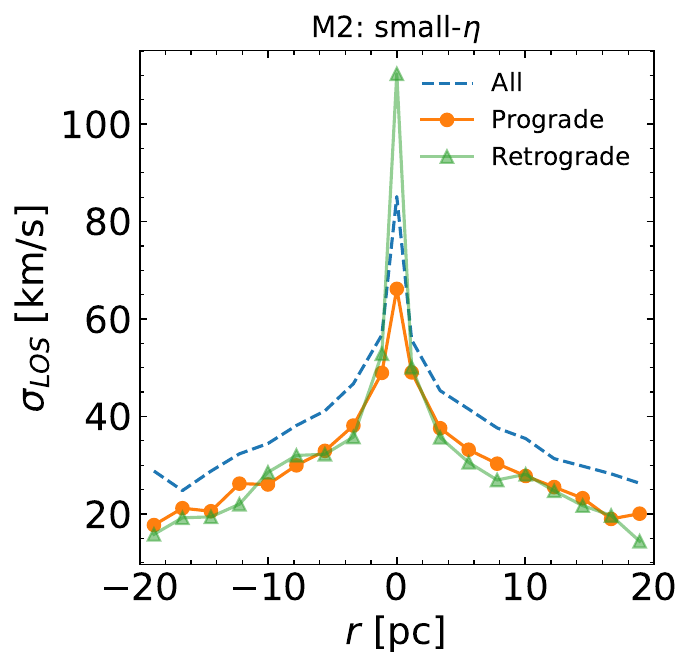}     
    \includegraphics[width=0.22\textwidth]{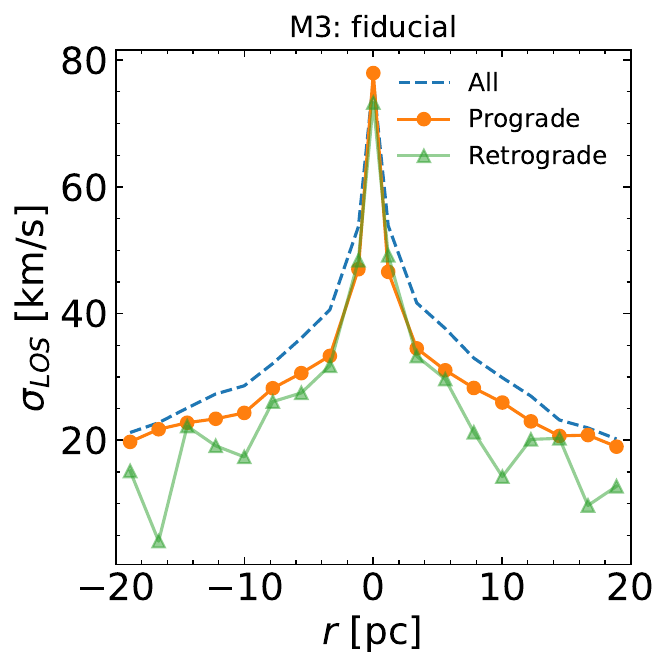}    
    \includegraphics[width=0.22\textwidth]{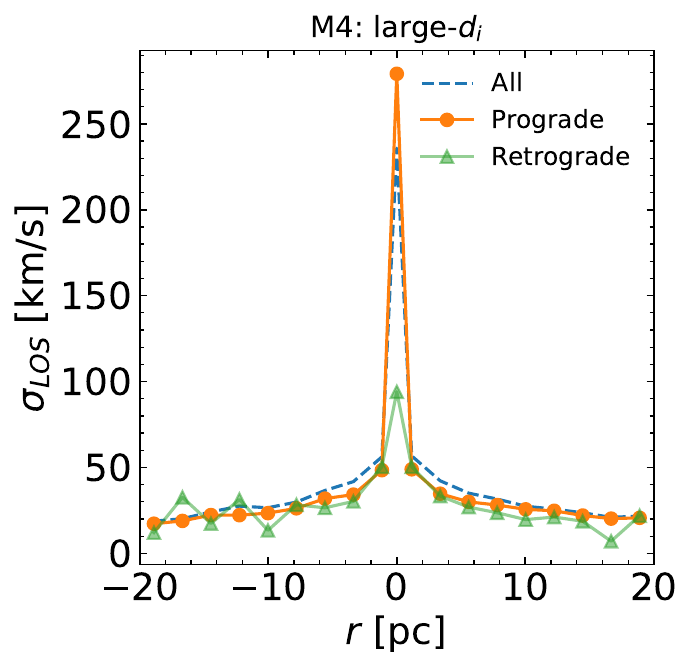}
    \includegraphics[width=0.22\textwidth]{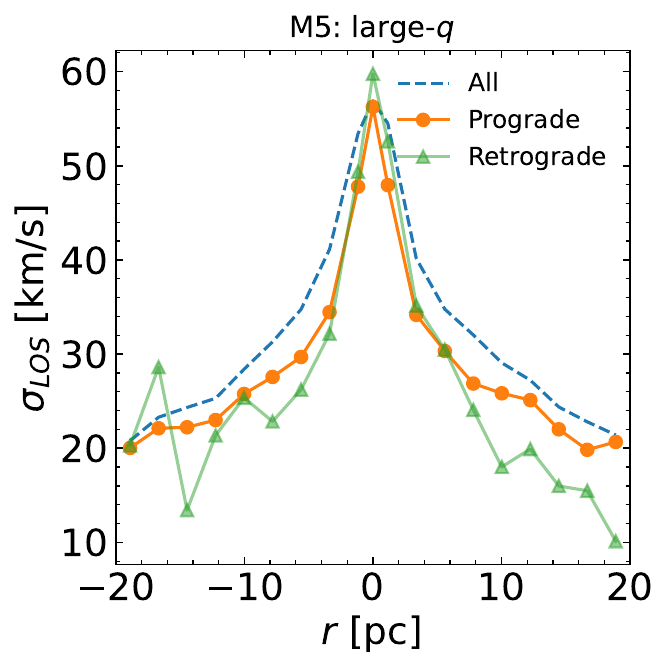}
    \includegraphics[width=0.22\textwidth]{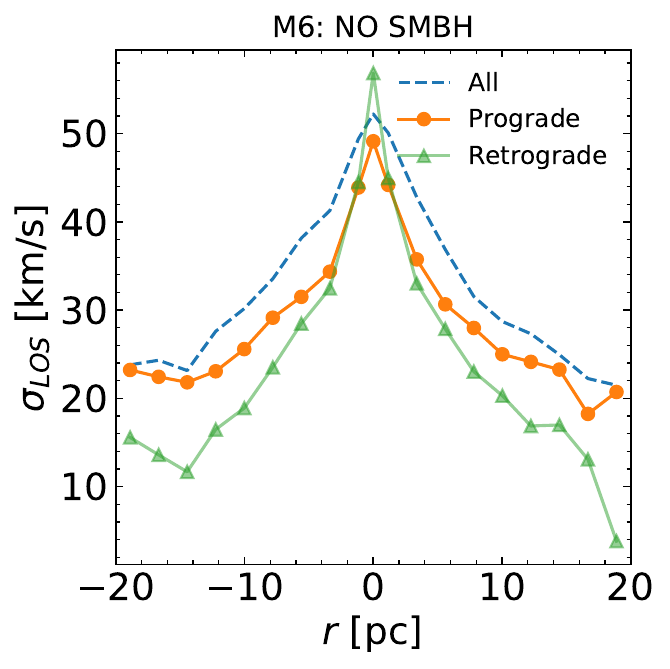}
    \includegraphics[width=0.235\textwidth]{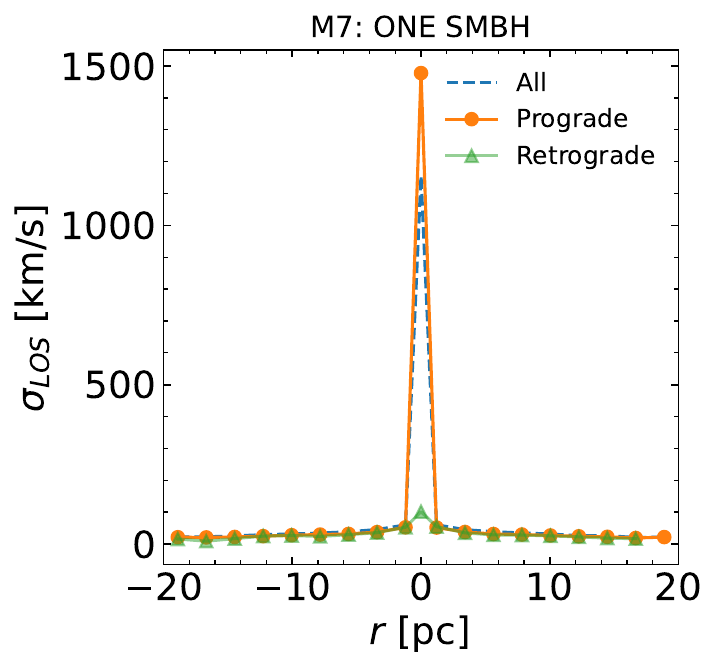}

    \caption{Line-of-sight velocity dispersion of the entire NSC (dashed blue line), of the prograde population (orange line with bullets), and of the retrograde population (green line with triangles). The behaviour of the two populations, and the resulting velocity dispersion of the entire cluster, vary depending on the initial merger configuration. }
    \label{fig:dispersion_curves}
\end{figure*}

\begin{figure*}
\raggedright
    \includegraphics[width=0.22\textwidth]{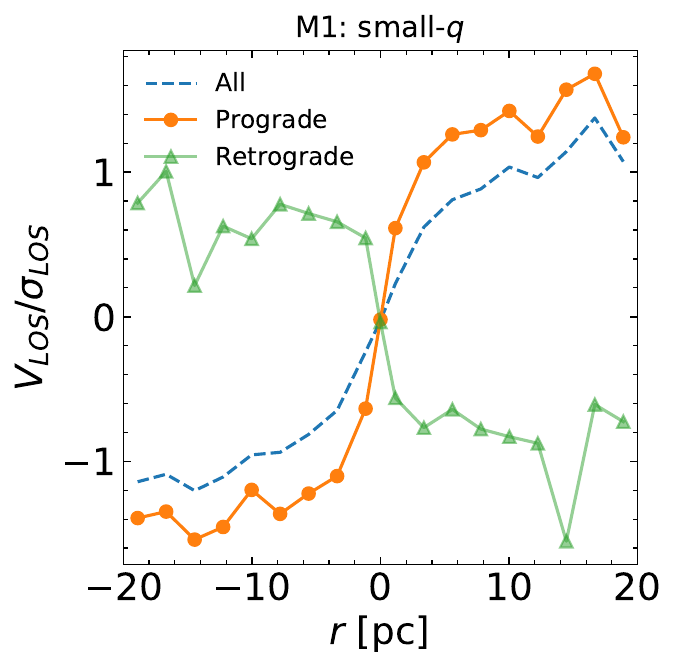}   
    \includegraphics[width=0.22\textwidth]{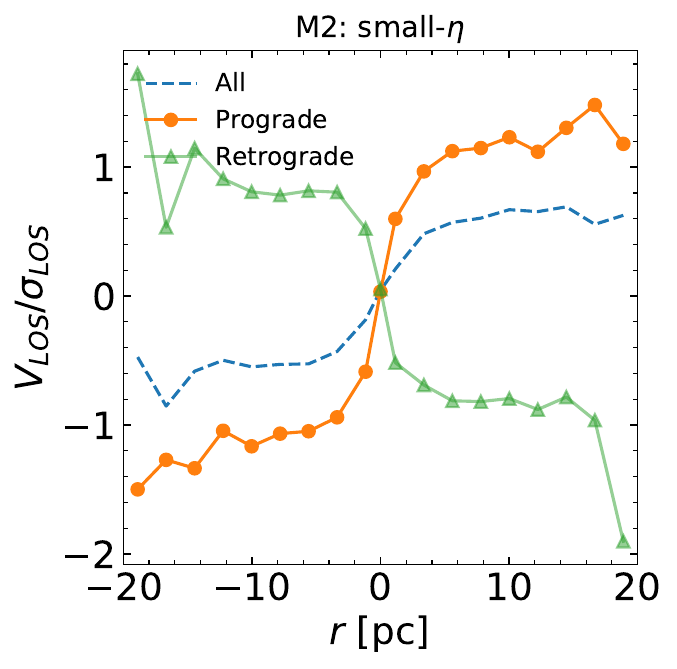}     
    \includegraphics[width=0.22\textwidth]{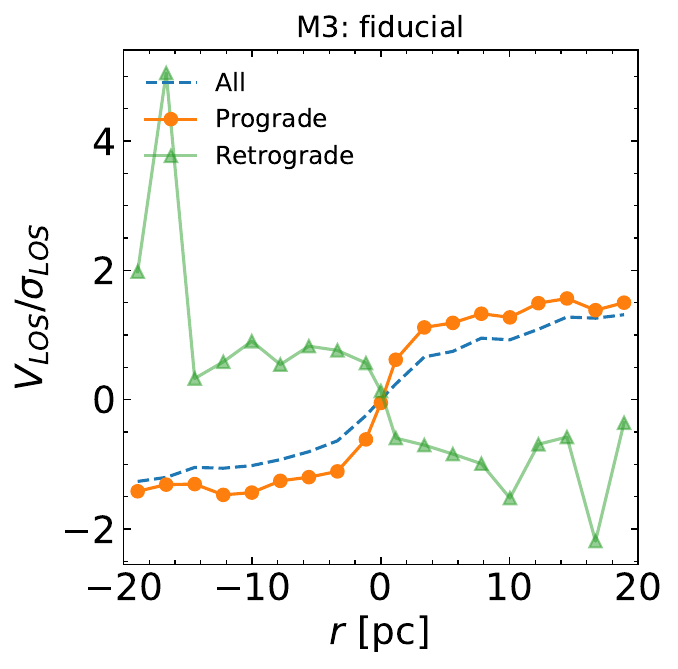}    
    \includegraphics[width=0.22\textwidth]{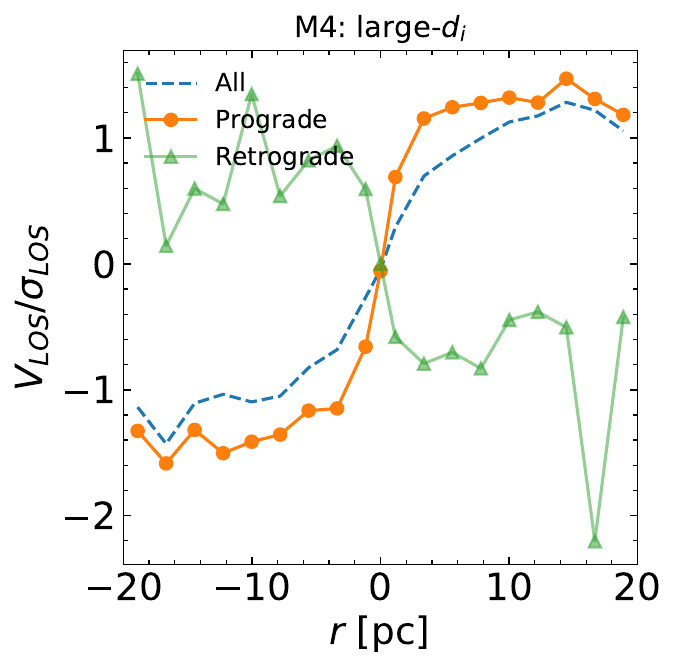}
    \includegraphics[width=0.22\textwidth]{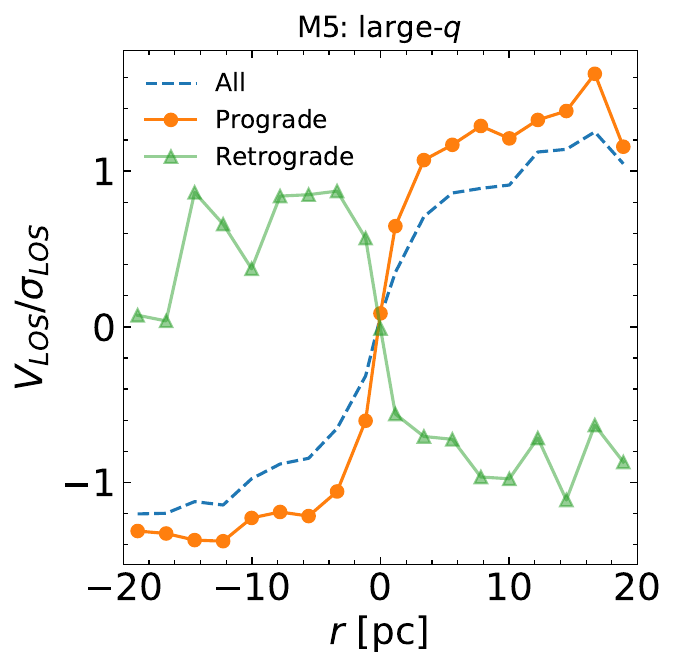}
    \includegraphics[width=0.22\textwidth]{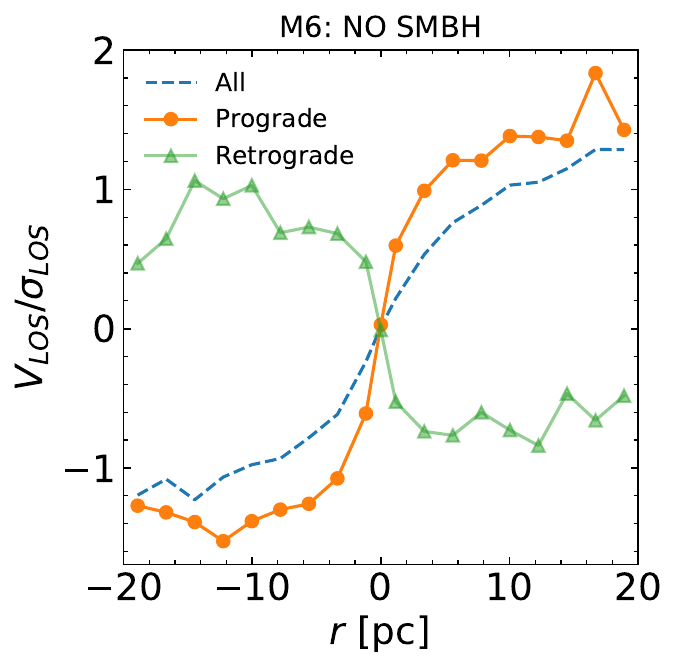}
    \includegraphics[width=0.22\textwidth]{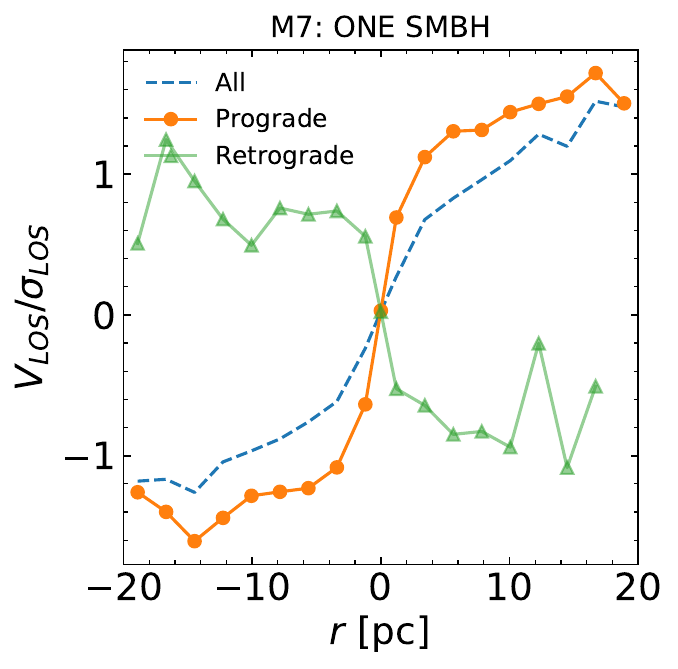}

    \caption{$V_{LOS}/\sigma_{LOS}$ ratio for the entire NSC (dashed blue line), of the prograde population (orange line with bullets), and of the retrograde population (green line with triangles). This curves confirm that the prograde population is more rotationally supported than the retrograde one.}
    \label{fig:Vsigma}
\end{figure*}

\begin{figure*}
\raggedright
    \includegraphics[width=0.22\textwidth]{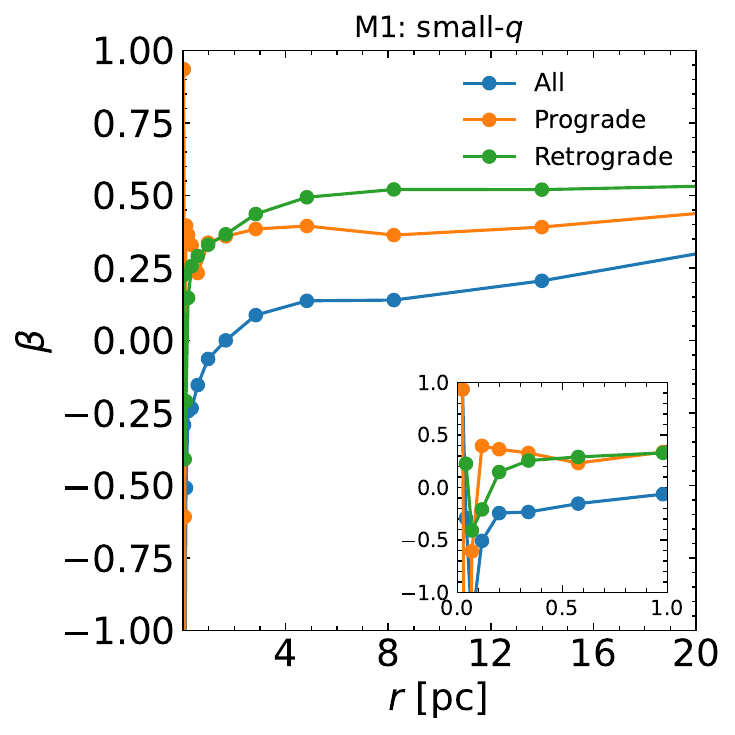}   
    \includegraphics[width=0.22\textwidth]{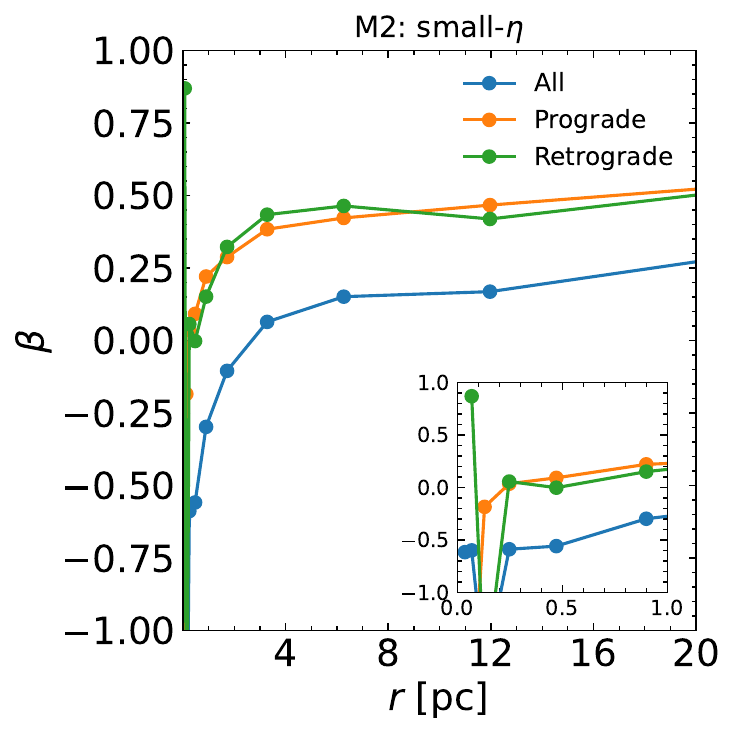}     
    \includegraphics[width=0.22\textwidth]{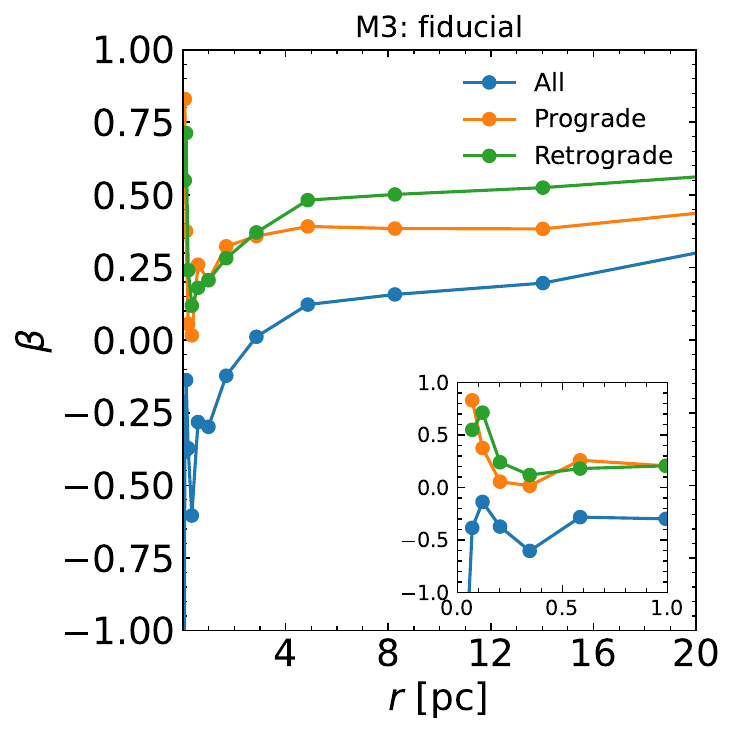}    
    \includegraphics[width=0.22\textwidth]{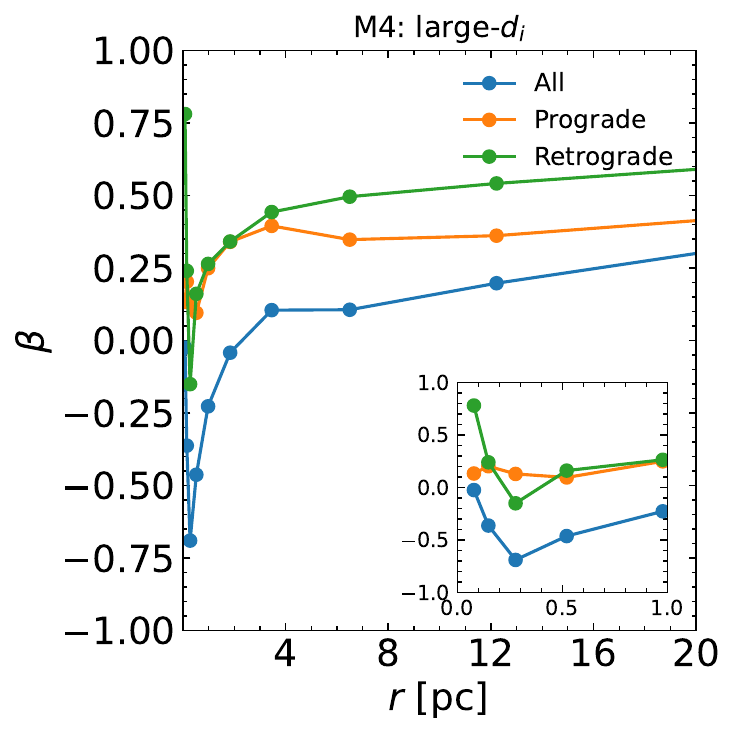}
    \includegraphics[width=0.22\textwidth]{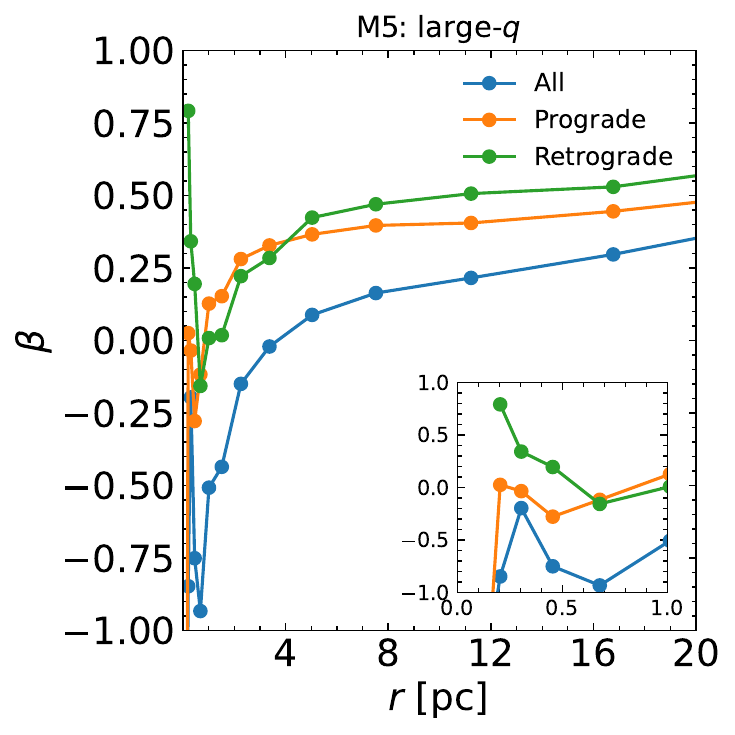}
    \includegraphics[width=0.22\textwidth]{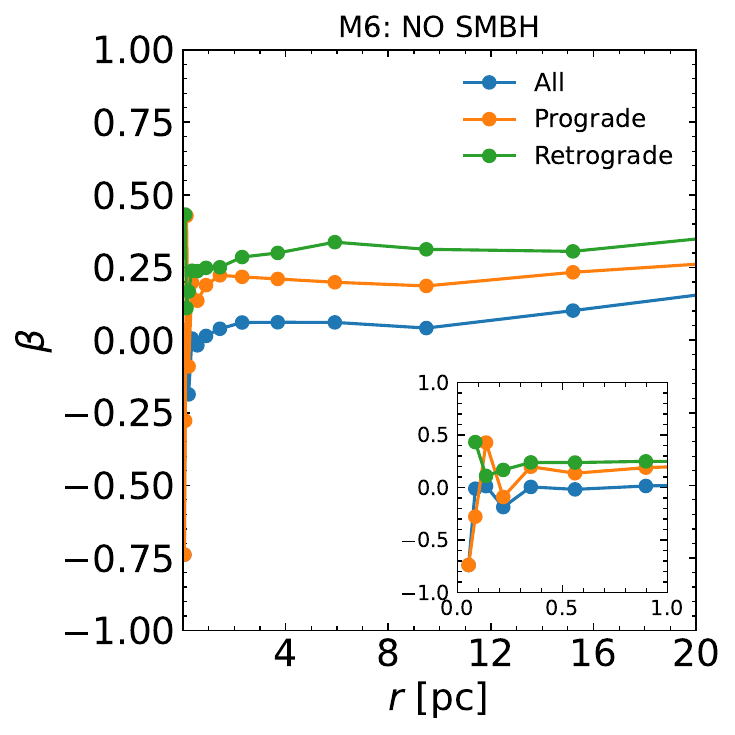}
    \includegraphics[width=0.22\textwidth]{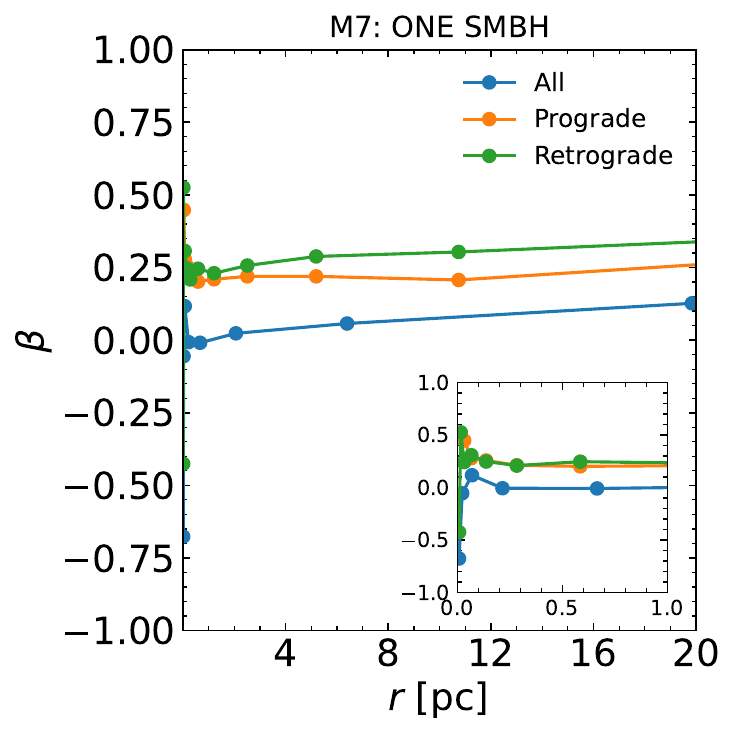}

    \caption{Velocity anisotropy parameter $\beta$ for the entire cluster (blue) as well as for the prograde (orange) and retrograde (green) populations. The behaviour of $\beta$ depends on the initial conditions adopted for each model. The inset in each panel shows the behaviour of $\beta$ in the cluster central regions.}
    \label{fig:beta_curves}
\end{figure*}

\end{appendix}
\end{document}